\documentclass[12pt]{article}
\usepackage{multibox}

\makeatletter \@addtoreset{equation}{section} \makeatother

\def\theequation{\thesection.\arabic{equation}}

\def\a{\alpha}
\def\tga{{\tilde{\alpha}}}
\def\b{\beta}

\def\d{\delta}

\def\m{\um}

\def\o{\omega}
\def\s{\sigma}

\def\be{\begin{equation}}
\def\ee{\end{equation}}
\def\bqn{\begin{eqnarray}}
\def\eqn{\end{eqnarray}}

\def\cc{{\cal C}}

\def\ce{{\cal E}}

\def\Box{{
\begin{picture}(7,9)
{\linethickness{.250mm}
\put(00,00){\line(1,0){7}}%
\put(00,7){\line(1,0){7}}%
\put(00,0){\line(0,1){7}}%
\put(07,0){\line(0,1){7}}%
}
\end{picture}   \,
}}

\def\cl{{\cal L}}

\def\car{{\cal R}}

\def\be{\begin{equation}}
\def\ee{\end{equation}}
\def\bea{\begin{eqnarray}}
\def\eea{\end{eqnarray}}

\newcommand{\D}{{\tilde{D}}}

\begin{document}

\begin{flushright}
{hep-th/0504090}\\
\vspace{1mm}
 FIAN/TD/07--05\\
\vspace{-1mm}
\end{flushright}\vspace{1cm}

\begin{center}
{\large\bf Actions, Charges and Off-Shell Fields in the\\
\vskip0.2cm
 Unfolded Dynamics Approach}
 \vglue 0.6  true cm \vskip0.8cm {M.A.~Vasiliev}
\vglue 0.3  true cm

I.E.Tamm Department of Theoretical Physics, Lebedev Physical Institute,\\
Leninsky prospect 53, 119991, Moscow, Russia
\vskip1.3cm
\end{center}

\begin{abstract}
Within unfolded dynamics approach, we represent
 actions and conserved charges as elements of
cohomology of the  $L_\infty$ algebra
underlying the unfolded formulation of a given dynamical system.
The unfolded off-shell constraints
for symmetric fields of all spins in Minkowski space are shown to
have the form of zero curvature  and covariant
constancy  conditions
for 1-forms and 0-forms taking values in an appropriate
star product algebra.
Unfolded formulation of Yang-Mills and Einstein equations is
presented in a closed form.
\end{abstract}

\newcommand{\xv}{{X(p,y)\Big|_{V_-}}}
\newcommand{\xm}{{X(p,y)\Big|_{-}}}
\newcommand{\ty}{\hat{y}}
\newcommand{\bee}{\begin{eqnarray}}
\newcommand{\eee}{\end{eqnarray}}
\newcommand{\nn}{\nonumber}
\newcommand{\lis}{Fort1,FV1,LV}
\newcommand{\hy}{\hat{y}}
\newcommand{\by}{\bar{y}}
\newcommand{\bz}{\bar{z}}
\newcommand{\go}{\omega}
\newcommand{\e}{\epsilon}
\newcommand{\f}{\frac}
\newcommand{\p}{\partial}
\newcommand{\BB}{F}
\newcommand{\half}{\frac{1}{2}}
\newcommand{\ga}{\alpha}
\newcommand{\gal}{\alpha}
\newcommand{\U}{\Upsilon}
\newcommand{\C}{{\bf C}}
\newcommand{\ups}{\upsilon}
\newcommand{\bu}{\bar{\upsilon}}
\newcommand{\dga}{{\dot{\alpha}}}
\newcommand{\dgb}{{\dot{\beta}}}
\newcommand{\gb}{\beta}
\newcommand{\gga}{\gamma}
\newcommand{\gd}{\delta}
\newcommand{\gl}{\lambda}
\newcommand{\gk}{\kappa}
\newcommand{\gep}{\epsilon}
\newcommand{\gvep}{\varepsilon}
\newcommand{\gs}{\sigma}
\newcommand{\V}{|0\rangle}
\newcommand{\ws}{\wedge\star\,}
\newcommand{\gee}{\epsilon}
\newcommand{\ggg}{\gamma}
\newcommand\ul{\underline}
\newcommand\un{{\underline{n}}}
\newcommand\ull{{\underline{l}}}
\newcommand\um{{\underline{m}}}
\newcommand\ur{{\underline{r}}}
\newcommand\us{{\underline{s}}}
\newcommand\up{{\underline{p}}}

\newcommand\hp{{\hbar}}
\newcommand\uq{{\underline{q}}}
\newcommand\ri{{\cal R}}
\newcommand\punc{\multiput(134,25)(15,0){5}{\line(1,0){3}}}
\newcommand\runc{\multiput(149,40)(15,0){4}{\line(1,0){3}}}
\newcommand\tunc{\multiput(164,55)(15,0){3}{\line(1,0){3}}}
\newcommand\yunc{\multiput(179,70)(15,0){2}{\line(1,0){3}}}
\newcommand\uunc{\multiput(194,85)(15,0){1}{\line(1,0){3}}}
\newcommand\aunc{\multiput(-75,15)(0,15){1}{\line(0,1){3}}}
\newcommand\sunc{\multiput(-60,15)(0,15){2}{\line(0,1){3}}}
\newcommand\dunc{\multiput(-45,15)(0,15){3}{\line(0,1){3}}}
\newcommand\func{\multiput(-30,15)(0,15){4}{\line(0,1){3}}}
\newcommand\gunc{\multiput(-15,15)(0,15){5}{\line(0,1){3}}}
\newcommand\hunc{\multiput(0,15)(0,15){6}{\line(0,1){3}}}
\newcommand\ls{\!\!\!\!\!\!\!}

\section{Introduction}

Nonlinear dynamics of massless higher spin
(HS) gauge fields was formulated
\cite{more,non} within the unfolded formulation \cite{unf}
in which the dynamical equations
have a form of certain covariant constancy and zero-curvature conditions
with respect to space-time coordinates. This form of the equations
is useful in various respects because its formal consistency
controls gauge symmetries and diffeomorphisms. The unfolded
formalism works beautifully and efficiently for the infinite sets of
fields of all spins through the concise formalism of certain
generating functions \cite{more,non}. Although the unfolded
formulation exists, in principle, for any dynamical system,
it may not necessarily be clear how to develop it
in a closed form for one or another
specific model. In particular, the unfolded
formulation of the nonlinear Einstein and Yang-Mills  equations was
not known so far, although some first order corrections have been
explicitly found in \cite{tri} for pure Einstein
gravity. One of the motivations of this paper was to fill in this gap by
constructing unfolded formulation for these nonlinear lower spin systems.

One has to distinguish between off-shell
unfolded formulation that only takes into account the Bianchi
identities and on-shell formulation that takes into
account field equations. The formulation of the HS dynamics
in $AdS_d$ developed in \cite{non} gives both the off-shell formulation
(as was emphasized in \cite{SSS}) and on-shell formulation. The latter
results from  factorization of the ideal associated with the
off-shell degrees of freedom that are trivial on-mass-shell
\cite{non} (for more details see \cite{solv}). As shown originally
in \cite{FV1}, consistent HS interactions with gravity require
$AdS$ background  with non-zero cosmological constant as the most
symmetric vacuum. This is why the formalism of
\cite{non} was designed to incorporate
$AdS$ background and $AdS$ symmetries in a most natural way.

The relevance of the $AdS$ background concerns
the action principle and, therefore, HS field equations.
There is no reason, however, to expect that  $AdS$ description
may play any significant role in the off-shell formulation. Indeed,
Bianchi identities in any system are merely identities: no
obstruction can result from rewriting them in the unfolded
form in any background. Therefore, the unfolded
off-shell formulation of \cite{non} should be expected
to admit a flat limit in one or another way.
A new result presented in this paper is the unfolded
form of off-shell nonlinear constraints for symmetric fields
of all spins in flat background.

So far the unfolded formulation was applied \cite{unf}
(see also \cite{Gol,s3,solv} for reviews)
to the analysis of the field equations and/or constraints.
In this paper we extend this approach to the action level.
The explanation of the general idea and main ingredients
of the unfolded formulation is the content of section
\ref{Unfolded dynamics}. Let us note that a closely
related structure is $L_\infty$ algebra \cite{Linf}.

The unfolded form of the off-shell constraints for symmetric
fields turns out to be so simple and inspiring that we
start  in section \ref{Summ} with giving
the final results leaving detailed
proofs and explanations for the rest of the paper which is organized
as follows. After recalling unfolded formulation of free
symmetric fields in section \ref{Free Fields Unfolded} and
 a few relevant algebraic properties
in section \ref{$sp(2)$ generators} we show in section
\ref{Linearization} that linearization of the nonlinear equations
of section \ref{Summ} correctly reproduces the linearized
unfolded constraints for massless fields. Nonlinear unfolded
constraints are analysed in section
\ref{Nonlinear off-mass-shell unfolding}. In particular,
in subsection \ref{sleq2} the unfolded formulation of the off-shell
constraints for Yang-Mills and Einstein theories
is obtained in the closed form. The on-shell unfolded formulation
for Yang-Mills and Einstein equations is discussed in section
\ref{field equations for lower spins}. It is shown here that the
key constraints still have the form of Yang-Mills and
Einstein equations although in terms of different variables.
Alternative versions of unfolded constraints are discussed in section
\ref{contr}.  In Appendix A we extend the construction of unfolded
dynamics to the general case with explicit dependence on space-time
coordinates. Theorems on $\sigma_-$ cohomology, which play
central role in the dynamical analysis of an unfolded system,  are
reproduced in Appendix B to make this paper as self contained as possible.

\section{Unfolded Dynamics}
\label{Unfolded dynamics}

\subsection{Unfolded equations}
\label{fda}

Let $M^d$ be a $d$-dimensional space-time manifold
with coordinates $x^\un$ ($\un = 0,1,\ldots d-1$).
By unfolded formulation of a linear or nonlinear
system of differential equations and/or constraints
in $M^d$ we mean its
equivalent reformulation in the first-order form
\be
\label{unf} dW^\ga (x)= G^\ga (W(x))\,,
\ee
where $ d=dx^\un  \frac{\p}{\p x^\un}\, $
is the de Rham differential in $M^d$, $W^\ga(x)$
is a set of degree $p_\ga$ differential forms
and $G^\ga (W)$ is some degree $p_\ga +1$
function of the differential forms $W^\ga$
\be
G^\ga (W) =
\sum_{n=1}^\infty f^\ga{}_{\gb_1\ldots \gb_n} W^{\gb_1}\wedge \ldots
\wedge W^{\gb_n}\,,
\ee
where the coefficients $f^\ga{}_{\gb_1\ldots
\gb_n}$ satisfy the (anti)symmetry condition
\be
f^\ga{}_{\gb_1\ldots\gb_k\ldots\gb_l \ldots \gb_n} =
(-1)^{p_{\gb_k}p_{\gb_l}}
f^\ga{}_{\gb_1\ldots\gb_l\ldots\gb_k \ldots  \gb_n} \,
\ee
(an extension to the supersymmetric case
with addition boson-fermion grading is straightforward)
and $G^\ga$ satisfies the  condition
\be \label{BI} G^\gb (W)\wedge \f{\p
G^\ga (W)} {\p W^\gb}  =0\,
\ee
equivalent to the
following generalized Jacobi identity on the structure coefficients
\be \label{jid}
\sum_{n=0}^{m} (n+1) f^\gamma{}_{[\b_1 \ldots
\b_{m-n}}  f^\a{}_{\gamma\b_{m-n+1} \ldots \b_m\}} =0\,,
\ee
where
the brackets $[\,\}$ denote an appropriate (anti)symmetrization of
all indices $\b_i$. Strictly speaking, the generalized Jacobi
identities (\ref{jid}) have to be satisfied only at $p_{\a} < d$ for
the case of a $d$-dimensional manifold ${\cal M}^d$ where any
$d+1$-form is zero. Given solution of (\ref{jid}) it defines a
 free differential algebra.
We  call a free differential algebra {\it
universal} if the generalized Jacobi identity is true for all values
of indices, {\it i.e.,} independently of a particular value of
space-time dimension. The HS free differential algebras discussed in
this paper belong to the universal class.  Note that every
universal free differential algebra is associated with
some $L_\infty$ algebra\footnote{The difference is that
a form degree $p_\a$ of $W^\a$ is fixed in a universal
free differential algebra
while $W^\a$ in $L_\infty$  are treated as local coordinates of
a graded manifold. A universal free differential algebra
can be obtained from a $L_\infty$ algebra by an
appropriate projection to specific form degrees.
(For more detail on this relationship we refer the reader
to \cite{BG}. I am grateful to Maxim Grigoriev
for illuminating discussion of this relationship.)
The language of  free differential algebras used in
 the earlier papers \cite{unf,s3} (however, with the
important extension to the case with 0-forms included),
where the unfolding approach was originally suggested
and applied to the analysis of HS models,
was introduced into field-theoretical literature in
\cite{FDA}. In the absence of $0$-forms, the structure of
these algebras was classified by Sullivan \cite{Sullivan}.}
\cite{Linf}.

The condition (\ref{BI}),  which can equivalently be rewritten as
\be
\label{qdif}
Q^2 =0\,,\qquad Q= G^\ga (W)  \f{\p}{\p W^\ga}\,,
\ee
guarantees the formal consistency of the unfolded system (\ref{unf})
with $d^2=0$ for all $W^\ga$. This form of the compatibility condition
manifests close relation of our construction with $Q$-manifolds
\cite{AKSZ} and, more generally, with the Batalin-Vilkovisky
approach \cite{BV} (see \cite{BGST} for a recent discussion
of the latter relationship). The unfolded equations (\ref{unf})
now imply
\be
\label{unf1}
d F(W(x)) =  Q (F(W(x))
\ee
for any function $F(W)$ that does not contain explicit dependence
on the coordinates $x^\un$, i.e., depends only on $W^\a$.
This form of the unfolded equations is analogous to the Hamiltonian
equations in the  standard $1d$ Hamiltonian dynamics.
(For completeness, let us mention that, as explained in more detail
in Appendix A, the unfolded formulation
admits a natural extension to the case where the function
$G^\ga=G^\ga (W,x)$ depends explicitly on the coordinates $x^\un$.)

The equation (\ref{unf}) is invariant under the gauge transformation
\be \label{delw} \delta W^\ga = d \varepsilon^\ga +\varepsilon^\gb
\frac{\p G^\ga (W) }{\p W^\gb}\,,
\ee
where the derivative $\frac{\p }{\p W^\gb}$ is left and
the gauge parameter
$\varepsilon^\ga (x) $ is a $(p_\ga -1)$-form.  ($0$-forms
$W^\a$ do not give rise to gauge parameters.)

An example of a free differential algebra
can be constructed as follows. Let $h$ be a Lie algebra, a basis of
which is the set $\{T_\a\}$. Let $\o=\o^\a T_\a$ be a 1-form taking
values in $h$. If one chooses $G (\,\o)=-\o\wedge\o\equiv -\frac{1}{2}
\o^\a \wedge\o^\b [T_\a , T_\b]$, then the equation (\ref{unf}) with
$W=\o$ is the zero-curvature equation
\be
\label{0cur} d\o+\o\wedge
\o=0\,.
\ee The relation (\ref{BI}) amounts to the usual Jacobi
identity for the Lie algebra $h$. (\ref{delw}) is then
the usual gauge transformation of the connection $\o$.

If the set $W^\a$ also contains some $p$-forms denoted by $\cc^i$
({\it e.g.} $0$-forms) and if the functions $G^i$ are linear in $\o$
and $C$
\be
\label{lin}
 G^i =- \o^\a(T_\a)^i {}_j \wedge \cc^j\,,
\ee
 then the relation
(\ref{BI}) implies that the matrices $(T_\a)^i {}_j$ form some
representation $T$ of $h$, acting in a space $V$ where the $\cc^i$
take their values. The corresponding equation (\ref{unf}) is a
covariant constancy condition \be \label{covc} D_\o \cc=0 \ee
 with $D_\o\equiv d+\o$ being a covariant derivative
in the $h$-module $V$.

The zero-curvature equations (\ref{0cur}) usually describe
background geometry in a coordinate independent way.
For example, let $h$ be the Poincare algebra with the gauge fields
\be
\o (x)=e^n (x) P_n
+\o^{nm} (x) L_{nm}\,,
\ee
where $P_n$ and $L_{nm}$ are generators of
translations and Lorentz transformations with the respective gauge
fields $e^n (x)$ and $\o^{nm}(x)$ to be identified with the frame 1-form
and Lorentz connection, respectively
(fiber Lorentz vector indices $m,n\ldots$ run from 0 to $d-1$
and are raised and lowered by the flat Minkowski metric).
It is well-known that the zero-curvature
condition (\ref{0cur}) for the Poincar\`{e} algebra amounts to the
zero-torsion condition
\be
\label{tor}
R^n=de^n -\o^n{}_m \wedge e^m = 0\,,
\ee
which expresses $\o^n{}_m$ in terms of derivatives of
the frame field,
and the Riemann tensor vanishing condition
\be
\label{riem}
R^{mn}=d\o^{mn}
-\o^m{}_k \wedge \o^{kn}=0\,,
\ee
which implies flat Minkowski space-time geometry.

As a result, at the condition that
$e_{0\,\un}{}^n (x)$ is a nondegenerate matrix, the zero-curvature
condition (\ref{0cur}) for the Poincar\`{e} algebra describes
Minkowski space-time in a coordinate independent way.
By choosing a different Lie algebra $h$ one can describe
a different background like, e.g., (anti-) de Sitter.
The covariant constancy
equation (\ref{covc}) can describe linear equations in a chosen
background.

\subsection{Dynamical content}
\label{dyncon}

Any consistent system of partial
differential equations and/or constraints
can, in principle, be re-written in the unfolded form
by adding enough auxiliary variables \cite{Vasiliev:1992gr}.
The unfolded formulation is a multidimensional covariant
({\it i.e.} coordinate independent) generalization of the first--order
formulation
\be
dt \f{\p}{\p t}{q}^\ga = G^\ga (q)\,,\qquad G^\ga = e F^\ga
\ee
available for any system of ordinary differential
equations by adding auxiliary variables to be identified with
higher derivatives of the dynamical variables of the original
system of differential equations. Here $e$ is a einbein 1-form
that can be identified with $dt$ because a one-dimensional
space-time is always flat.
In this $d=1$ example,
the condition (\ref{BI}) trivializes because $e\wedge e=0$,
{\it i.e.} any function $F^\ga (q)$ is allowed. Note that
this condition is true for an arbitrary number of coordinates
of the ambient space (i.e., $dx^\um$),
provided that $e = dx^\um e_\um$ does not carry fiber
indices. This means that the system is universal.
Although, for the case of $d>1$ with $e^n$ having $d$ components,
nontrivial consistency conditions have to be
taken into account, the simplest $d=1$
example illustrates the general mechanism.

The structure of first-order ordinary differential
equations is as follows
\be
\label{eqex}
 \f{\p}{\p t}{q}^\tga_i =a_i^\tga {q}^\tga_{i+1} +\ldots\,,
\qquad i=0,1,2,\ldots\,,
\ee
where $a_i^\tga$ are some coefficients and dots denote
higher order nonlinear corrections. If all coefficients $a^\tga_i$
are different from zero,
the equations (\ref{eqex}), treated perturbatively, describe
a set of constraints that express all ${q}^\tga_{i+1}$ via
derivatives of  ${q}^\tga_{0}$. If some coefficient $a_j^\tga$
vanishes, this means that there is some nontrivial
differential equation on ${q}^\tga_{0}$, which is of
order $j$ at the linearized level.
In the first order formulation
(in particular, in the Hamiltonian formalism) the initial data
problem is fixed in terms of values of all variables $q^\ga$
at a given point of ``space-time" $t=t_0$.

In the general case of $d>1$ these properties have clear analogues.
Nontrivial dynamical fields ({\it i.e.,} those that are different from
{\it auxiliary fields} expressed via derivatives of the
dynamical fields), gauge symmetries and true differential field
equations, are classified in terms of the so-called $\sigma_-$
cohomology \cite{sigma}
 that roughly speaking controls zeros
among the coefficients analogous to $a^\tga_i$ of the linearized
equations.
The $\sigma_-$ cohomology is a perturbative
concept that emerges in the linearized analysis with
\be
W^\a (x) = W^\a_0 (x) + W_1^\a (x)\,,
\ee
where $W^\a_0 (x)$ is a particular solution
of (\ref{unf}) and $W_1^\a (x)$ is treated as a perturbation.
$W^\a_0 (x)$ is nonzero in a field-theoretical
system because it should describe a background gravitational
field. Typically, $W_0^\a (x)= (e_0^n (x)\,, \o_0^{nm} (x)),$
where $e_0^n (x)$ and $ \o_0^{nm} (x)$ are frame and Lorentz
connection 1-forms of the background space-time
(e.g., Minkowski, $(A)dS$, etc.) Linearized equations (\ref{unf})
\be
dW_1^\a (x) = W_1^\b (x) \f{\delta G^\a}{\delta W^\b }\Big |_{W=W_0}
\ee
can be rewritten in the form (\ref{covc})
\be
\label{d0}
D_0 W_1^\a (x)=0\,,
\ee
where $D_0$ is some differential which squares to zero as a
consequence of the consistency condition (\ref{BI}),
\be
\label{d02}
D_0^2 =0\,.
\ee

Usually a set of fields $W_1$ admits a grading $G$ with the
spectrum bounded from below. The grading $G$  typically counts a
rank of a tensor (equivalently, a power of an appropriate
generating polynomial). Suppose that \be \label{d0s} D_0 = \D_0
+\sigma_- +\sigma_+\,, \ee where \be \label{gs-} [G\,,\sigma_- ] =
- \sigma_-\,,\qquad [G\,,\D_0 ]= 0\, \ee and $\sigma_+$ is a sum
of operators of positive grade. From (\ref{d02}) it follows that
\be \sigma_-^2 = 0\,. \ee
Provided that $\sigma_-$ acts vertically ({\it i.e.} does not
differentiate $x^\un$), the dynamical content of the dynamical
system under investigation is determined by cohomology of
 $\sigma_-$. Namely, as shown in \cite{sigma}, for a
$p_\a$-form $W^\a$ that takes values in a vector space $V$,
$H^{p+1} (\sigma_- ,V)$, $H^{p} (\sigma_- ,V)$ and $H^{p-1}
(\sigma_- ,V)$ describe, respectively, differential equations,
dynamical fields and differential gauge symmetries encoded by the
equation (\ref{d0}).  (For more detail  see Appendix B and \cite{solv}.)
 The case with $H^{p+1} (\sigma_- ,V)=0$ is
analogous to that of (\ref{eqex}) with all
coefficients $a^\tga_i$ different from zero
where no differential equations  on the dynamical
variables are imposed. In this case,
 the equations (\ref{unf}) just express the Bianchi identities
for the constraints on auxiliary fields. Equations of this type
will be referred to as {\it off-shell}.
(Let  us stress that this definition is true both for
linear and for non-linear cases: nonlinear equations are
off-shell if their linearization is off-shell.)

The degrees of freedom, that is variables that fix a (local)
solution of the equations (\ref{unf}) modulo gauge ambiguity,
are values of all 0-forms $C^\phi(x)$ among $W^\ga (x)$,
taken at any given point $x^\un=x^\un_0$ \cite{s3}
analogous to $t_0$ of the $d=1$ case. For a field-theoretical
system with infinite number of degrees of freedom to be described
this way an infinite set of 0-forms has to be introduced. For a
topological system with a finite number of degrees of freedom,
unfolded equations with a finite number of 0-forms may be
available.

Whether a system is on-shell or off-shell depends not only on
$G^\ga (W)$ but, in first place, on a space-time $M^d$ and a
chosen vacuum solution $W_0^\a (x)$.
In particular, dynamical interpretation may be different for
different dimensions of the space-time.  On the other hand,
from the unfolded equations
(\ref{unf}) it follows that the dependence on the coordinates
$x^\un$ is reconstructed in terms of values of the fields
$W^\ga (x)$ at any given point $x_0^\un$. This means, in
turn, that the role of coordinates $x^\un$ is somewhat auxiliary
as was first noted in the context of HS
dynamics in \cite{algasp}. As emphasized in \cite{Mar}
the role of coordinates is that they help to visualize
physical local events via a specific physical process.
Moreover, a result of such a visualization
to large extent depends not on a chosen manifold $M^d$, where
one or another equation (\ref{unf}) is written, but mostly
on the specific form of the function $G^\a$. This property was
used for practical purposes of equivalent reformulation
of field equations in different space-times with
additional commuting  \cite{BHS,Mar} and anticommuting \cite{ESS}
coordinates.

\subsection{Actions and conserved charges}
\label{actions}
Independently of the dynamical interpretation of a
given unfolded system, its invariants like actions and conserved
charges turn out to be associated with the cohomology of the
differential $Q$ (\ref{qdif}).

First suppose that the system (\ref{unf}) is off-shell. Let us
extend the set of forms $W^\ga$ with  a $d-1$-form $E$ and a
$d$-form $L$ extending  the equations (\ref{unf}) with the
equations \be \label{EL} dE =L -\cl (W)\,, \ee \be \label{dl} dL =
0\,, \ee where $\cl(W)$ is some {\it Lagrangian function}
 of the variables $W^\ga$ which is
$Q$-closed
\be
\label{qcl}
Q\cl=0\,:\qquad G^\ga (W)  \f{\p}{\p W^\ga} \cl(W) =0\,.
\ee
The latter condition guarantees that the extended system remains
consistent, i.e. $Q^\prime = Q +(L- \cl)\f{\p}{\p E}$ satisfies
$(Q^\prime )^2=0$.

We define action $S$ as the integral of $L$ over a $d$-cycle $M^d$
\be \label{action} S=\int_{M^d} L\,, \ee
which, in turn, can be
embedded into a larger space. By (\ref{dl}), the result is
independent of local variations of this embedding.
The action $S$ (\ref{action}) is gauge invariant.
Actually, the gauge transformation (\ref{delw}) gives
for $E$ and $L$
\be
\delta E = d \epsilon_E +\epsilon_L - \epsilon^\ga
           \f{\p \cl (W)}{\p W^\ga}  \,,
\ee
\be
\label{dL}
\delta L = d\epsilon_L \,.
\ee
Assuming that $M^d$ has no boundary (or that the
field variables fall down fast enough at infinity)
 the action (\ref{action}) remains invariant under
the transformation (\ref{dL}).
Therefore, the action (\ref{action}) is invariant under
the full set of the
gauge transformations (\ref{delw}).

One can use the gauge
parameter $\epsilon_L $ to gauge fix $E$ to zero.
In this gauge, the action (\ref{action}) takes the form
\be \label{action1} S=\int_{M^d} \cl (W)\,.
\ee
Taking into account (\ref{qcl}) it is easy to see directly
that this action is invariant under
the gauge transformations (\ref{delw}).

If $\cl$ is $Q$-exact, {\it i.e.} $\cl =
G^\ga (W)  \f{\p}{\p W^\ga} \ce$,
by a field redefinition
\be
E\to E^\prime -\ce (W)\,,
\ee
this case is equivalent to that with $\cl=0$, {\it i.e.}
$Q$-exact Lagrangians $\cl(W)$ do not generate nontrivial actions
(as is also obvious from (\ref{unf1})).
Thus, nontrivial invariant actions are
in the one-to-one correspondence with
the $Q$ cohomology of the original off-shell system.
A Lagrangian function $\cl$ is a representative of
a $Q$ cohomology class.

If the system is on-shell and a representative
$\cl$ of the $Q$ cohomology
is a $p$-form, the same formula (\ref{action1})
describes a conserved charge as an integral over a
$p$--cycle $\Sigma$
\be
q = \int_{\Sigma} \cl\,.
\ee

Taking into account that whether a dynamical
system is on-shell or off-shell depends
on the choice of space-time manifold $M^d$ rather than
on $G^\ga (W)$, we conclude that there is no big difference in the
unfolded dynamics approach between actions and conserved charges associated
with the same $Q$-cohomology in seemingly different dynamical systems
described by the same operator $Q$.
It is interesting to study physical consequences of this
surprising identification.

Another point we would like to stress is that the $Q$--cohomology
describes full invariant actions and charges of the unfolded system
at hand rather than perturbative deformations of interactions, as one
would normally expect of a cohomology.

\subsection{Scalar field example}
\label{Free massless scalar}

To describe an off-shell scalar
field $C (x)$ in the unfolded form, we introduce
following \cite{sigma} the infinite set of
$0$-forms $C_{m_1\ldots m_n}(x)$ ($n=0,1,2,\ldots$), which are
completely symmetric tensors \be \label{tr} C_{m_1\ldots
m_n}=C_{\{m_1\ldots m_n\}}\,.
\ee
The off-shell ``unfolded" equations are
 \be \label{un0} d C_{m_1\ldots m_n } =e_0^k C_{m_1
\ldots m_n k}\,,\quad (n=0,1,\ldots)\,, \ee where we use
 Cartesian coordinates with vanishing Lorentz
connection that allows us to replace the Lorentz covariant
derivative $D_0^L$ by the exterior differential $d$.
This system is formally consistent because applying
$d$ on both sides of (\ref{un0}) does not lead to any new condition
as $e_0^k \wedge e_0^l=-e_0^l \wedge e_0^k$. This property
implies that the space $V$ of $0$-forms
$C_{m_1 \ldots m_n}$ spans some representation of the Poincar\'e
algebra $iso(d-1,1)$. In other words, $V$ is an infinite-dimensional
$iso(d-1,1)$-module\footnote{Strictly speaking, to apply the general
argument of subsection \ref{fda} one has to check that the equation
remains consistent for any flat connection in $iso(d-1,1)$. It is
not hard to see  that this is true indeed.}.

 Let us identify the scalar field $C (x)$ with $C_{m_1
\ldots m_n}(x)$ at $n=0$. Then the first two equations of the system
(\ref{un0}) read
\be
\label{fieq}
\partial_\un C =C_\un \,,\qquad
\partial_\un C_\um= C_{\um\un}\,,
\ee
where we have identified the world and tangent indices via
$(e_0)_\um^m=\delta_\um^m$. The first of these equations just tells
us that $C_\un$ is the first derivative of $C$. The second one tells
us that $C_{\un\um}$ is the second derivative of $C$.
 All other
equations in (\ref{un0}) express highest tensors in terms of the
higher-order derivatives \be \label{hder} C_{\um_1 \ldots \um_n}=
\partial_{\um_1} \ldots \partial_{\um_n}C \ee and impose no
conditions on $C$.
 {}From this formula it is clear that the meaning of the 0-forms
$C_{\un_1 \ldots \un_n}$ is that they form a basis in the space of
all derivatives of the dynamical field
$C(x)$, including the derivative of order zero which is the field
$C(x)$ itself. Thus, the system (\ref{un0}) is
off-shell: it forms an infinite set of constraints which
express all highest tensors in terms of derivatives of $C$
according to (\ref{hder}).

The above consideration is simplified by means of
introducing the auxiliary coordinate $y^n$ and the generating
function $$ C (y|x)=\sum_{n=0}^\infty \frac{1}{n\,!}C_{m_1 \ldots
m_n}(x) y^{m_1} \ldots y^{m_n} $$ with the convention that
$ C (0|x)=C(x)$.
The equations (\ref{un0}) then acquire the simple form
\be
\label{xu} \frac{\partial}{\partial
x^\um} C (y|x)=\delta_\um^m\,\frac{\partial}{\partial y^m} C
(y|x)\,.
\ee
{}From this realization one concludes that the
translation generators in the infinite-dimensional module $V$ of the
Poincar\'e algebra are realized as translations in the $y$--space,
{\it i.e.}
$P_n=-\frac{\partial}{\partial y^n}\,.$ The equation
(\ref{xu}) reads as a covariant constancy condition
\be
dC(y|x)+e_0^n P_nC (y|x)=0\,\label{xu2}\,.
\ee

Comparing this formula with (\ref{d0s}) and (\ref{gs-}) we see
that the operator $-\sigma_-$ is the de Rham differential in the
$y$ space
\be
\sigma_- =- dx^n \frac{\partial}{\partial y^n}\,.
\ee
The grading operator
\be
G=y^n\frac{\partial}{\partial y^n}\,,\qquad [G\,, \sigma_- ]=-\sigma_-
\ee
counts a degree of
the variables $y^n$, that is a
rank $n$ of a tensor $C_{m_1\ldots m_n }$.
By the Poincar{\`e} lemma we see that
$H^0(\sigma_-, V)$ is one-dimensional that corresponds to
a single dynamical field $C(x)$, and $H^1(\sigma_-, V)=0$
that corresponds to the absence of dynamical equations on $C(x)$.

To put the system on-shell by imposing massless or massive
Klein-Gordon equation in flat Minkowski space
\be
\label{KG}
(\Box_x +m^2 ) C(x)=0\,,\qquad\Box_x = \frac{\partial^2}{\partial x^m
\partial x_m}
\ee
is equivalent to imposing the on-shell condition
in the fiber variables
 $$
(\Box_y +m^2)C (y|x)=0\,,\qquad\Box_y \equiv
\frac{\partial^2}{\partial y^m \partial y_m}\,.
$$ 
In terms of component tensors $C_{m_1\ldots m_n}(x)$
the latter conditions are equivalent
to certain tracelessness conditions which express first
traces of the fields in terms of the same set of fields.
In the massless case $m=0$, the condition is
\cite{Vasiliev:1992gr} that the
$C_{m_1\ldots m_n}(x)$ is traceless\footnote{Note that, as shown in
\cite{prok} for the $3d$ example and then in \cite{sigma}
for the general case,
even in the massive case,
one can still work with traceless tensors $C_{m_1\ldots m_n}(x)$
by modifying appropriately the form of the operator
$P_n$ in (\ref{xu2}).}. Indeed from (\ref{fieq}) it follows that
the tracelessness of $C_{nm}$ implies the Klein-Gordon equation.
We refer the reader to \cite{solv} to
see how the Klein-Gordon equation results from the $\sigma_-$
cohomology $H^1 (\sigma_- ,\hat{V})$ in the space $\hat{V}$ of
traceless tensors.

Note that in the $1d$ case, where indices
take only one value, the set of traceless
tensors $C_{m_1\ldots m_n }$ contains only two nonzero
components: those
with $n=0$ and $n=1$. These are the coordinate and momentum of
the standard first-order ({\it e.g.,} Hamiltonian) formalism.

Now we discuss invariant functionals in the scalar field case. Let
us start with the off-shell case. Consider \be \label{scall} \cl =
e^{n_1}\wedge \ldots \wedge e^{n_d} \epsilon_{n_1\ldots n_d } \ell
(C,C^n, C^{nm},\ldots )\,, \ee where $\epsilon_{n_1\ldots n_d }$ is
completely antisymmetric tensor and $\ell (C,C^n, C^{nm},\ldots )$
is an arbitrary Lorentz invariant function ({\it i.e.,} all vector
indices are contracted by the Minkowski metric). The differential
$Q$ acts as follows
\be Q C(y|x) = \Big (\o_{nm}y^n \f{\partial
}{\partial y_m} + e^m \f{\partial }{\partial y_m}\Big ) C(y|x)\,,
\ee
\be Qe^n =\o^{nm} \wedge e_m\,,\qquad Q\o^{nm}=\o^{nk}\wedge
\o_{k}{}^m\,,
\ee
where the 1-forms $\o_{nm}$ and $e^m$ describe
Lorentz connection and frame field in any coordinates. We see that
$Q\cl =0$ because (i) $\ell (C,C^n, C^{nm},\ldots )$ is Lorentz
invariant (which means that all terms with $\o_{nm}$ cancel out) and
(ii) \be e^{n_1}\wedge \ldots \wedge e^{n_d} \wedge e^m \equiv 0\,
\ee because antisymmetrization over $d+1$ indices $n$, which take
$d$ values, gives zero.

The formula (\ref{scall}) gives a most general
Lorentz invariant Lagrangian of a scalar field.
($Q$-cohomology describes actions because
$Q$-exact Lagrangians are total derivatives by the
unfolded equation (\ref{unf1}).) Let us note that
we arrived at Poincar{\` e} invariant Lagrangians
because the 1-forms in the model are
the gauge fields of the Poincar{\` e} algebra. Actually,
from the general consideration
of subsection \ref{actions} it follows that
the constructed action is invariant under localized
 Poincare gauge transformations. As explained in more detail
{\it e.g.} in \cite{solv}, assuming that the gravitational
 fields $\o^{nm}$ and $e^n$ are some fixed background fields
 this local symmetry reduces to the global Poincare symmetry
(in any chosen coordinate system).

Let us now discuss conserved charges for the on-shell scalar
field. We set
\be
\label{scallc}
\cl = \epsilon_{n_1\ldots n_d } e^{n_1}\wedge \ldots \wedge e^{n_{d-1}}
 J^{n_d} (C,C^n, C^{nm},\ldots )\,,
\ee
where $J^{k} (C )$  is a Lorentz vector. Using that
\be
\epsilon_{n_1\ldots n_{d-1} k }
e^{n_1}\wedge \ldots \wedge e^{n_{d-1}} \wedge e^m
 = d^{-1} \epsilon_{n_1\ldots n_{d}  }
\delta_{k}^m
e^{n_1}\wedge \ldots \wedge e^{n_{d}}
\,
\ee
we obtain
\be
Q \cl =(-1)^d d^{-1} e^{n_1}\wedge \ldots \wedge e^{n_{d}}
\epsilon_{n_1\ldots n_d } \sum_{k=1}^\infty
C_{m_1 \ldots m_k} \f{\delta J^{m_k}}{\delta
C_{m_1 \ldots m_{k-1}}}
\,.
\ee
For example, taking into account that $C_{n}{}^n =0$,
 the condition $Q\cl=0$ is satisfied by the standard
spin one current
\be
J_{n}^{\a\b} = C_n^\a C^\b - C_n^\b C^\a\,,
\ee
where $\a $ and $\b$ are color indices. One can analogously construct
all other conserved currents with higher spins\footnote{Note that,
in flat space, the
currents with manifest dependence on the coordinates
 can also be analysed by virtue of introducing coordinates as
new 0-form variables $x^n$ satisfying the unfolded equations
$d(x^n) -\o^{nm}x_m= e^n$.}
\cite{BBD,ans,Gol,KVZ}.
The $Q$ cohomology factors out so-called
improvements thus characterizing conserved charges
rather than currents.

The example of unfolded scalar field is so simple that one might think
the procedure is a sort of a trivial substitution of the original
$x$ coordinates with the fiber $y$ coordinates. This is true to
some extent at the free field level (especially in Minkowski
background space-time with no $x$-dependence of the coefficients
of the equations), but is not the case in less trivial
situations like those with  interactions.
The advantage of the formalism is that in the sector
of fiber variables it is always a sort of flat: in particular,
indices are contracted by the flat Minkowski metric tensor.
The situation here is
 analogous to that in the Fedosov quantization prescription \cite{fed}
which reduces the nontrivial problem of quantization in a curved
background to the
standard problem of quantization of the flat phase space, that, of course,
becomes an identity when the ambient space is flat itself. It is
worth to mention that this parallelism is not  accidental
because, as one can easily see, the Fedosov quantization prescription
provides a  particular case of the general unfolding approach
\cite{unf} in the dynamically empty ({\it i.e.} off-shell)
situation. Note that parallelism between unfolded dynamics and Fedosov
quantization was also  discussed recently in \cite{BGST}.

\section{Off-Shell Unfolded Fields in Minkowski Space}
\label{Summ}

Both off-shell and on-shell unfolded formulation of free
massless fields of all spins is by now well-known
(see, e.g. \cite{solv} for  a review and references).
Before recalling details of the unfolded free field theory in section
\ref{Free Fields Unfolded}, we formulate in this section
the final results for the full nonlinear
system of off-shell unfolded constraints for symmetric fields
of all spins in flat background and its lower spin
reductions to the Yang-Mills and Einstein theories.

To this end we introduce a 1-form $A(p,y|x)=dx^\un
A_\un (p,y|x)$ and 0-form $\BB (p,y|x)$ which depend on
the usual commuting space-time coordinates
$x^\un$ and a pair of fiber Lorentz vectors $p_n$ and $y^n$
(recall that both base indices $\um , \un\ldots$ and fiber Lorentz
indices $m,n\ldots$ run from 0 to $d-1$).
The variables $p_n$ and $y^n$ form a canonical pair
with nonzero commutation relations
\be
\label{py}
[p_m \,,y^n]_*  = \hp\delta^n_m\,,
\ee
where $[a\,,b]_*= a*b - b*a $, the deformation
``Planck constant" parameter
 $\hp$ is introduced for the future convenience and
the Weyl star product is defined in the standard way
\be
\label{weyl} (f*g)(p,y)
 = f(p,y)\exp{\f{\hp}{2}\Big (
\f{\overleftarrow{\p}}{{\p} p^n} \f{{\p}}{{\p} y_n} -
\f{\overleftarrow{\p}}{{\p} y^n} \f{{\p}}{{\p} p_n}\Big )}\,g(p,y)\,.
\ee

Let $D$ be the covariant derivative
\be D= \hp \,d+A= dx^\un \left (\hp
\f{\p}{\p x^\un} +A_\un (p,y|x)\right)\,.
\ee
The nonlinear unfolded off-shell HS formulation in flat space takes
extremely simple form of the zero-curvature and
covariant constancy  conditions
\be \label{d2}
D^2 =0 \,,\qquad D(\BB)=0
\ee
that is
\be
 dA +\hp^{-1} A*A =0 \,,\qquad d\BB +
 \hp^{-1} [A,\BB]_* =0\,.
\ee
In
addition there is a ``boundary condition"
that
\be
\label{bound}
A=A_0 +A_1\,,\qquad \BB= \BB_0 +\BB_1\,,
\ee
where
\be
\label{A0}
 A_0 = dx^\un {e_{\un}{}^n p_n
+\go_{\un}{}^{nm}p_n y_m} \,, \ee
and
\be \label{b0} \BB_0 = \half p_n
p^n
\ee
are zero-order parts while $A_1$ and $\BB_1$ are fluctuations.
Here $e_{\un}{}^n$ and
$\go_{\un}{}^{nm}=-\go_{\un}{}^{mn}$ describe frame 1-form
 and
Lorentz connection of the background Minkowski space-time, i.e.
 $e_{\un}{}^n (x)$ is a nondegenerate matrix and the vacuum
equation
\be
\label{da0}
dA_0 +\hp^{-1} A_0  * A_0 =0
\ee
is satisfied, which in turn is
equivalent to the zero-torsion condition (\ref{tor})
(the part proportional to $p^n$) and vanishing Riemann tensor
condition (\ref{riem}) (the part proportional
to $p^{[n}y^{m]}$). A solution that describes Minkowski
space-time in Cartesian coordinates is $e_{\un}{}^m
=\delta_\un^m$ and $\go_{\un}{}^{mn}=0$.

The fields $A(p,y|x)$ and $\BB(p,y|x)$ are power series in the
fiber variables $p_n$ and $y^m$.
It is convenient to expand them in powers of the
fiber momenta $p_n$
\be
A(p,y|x)=\sum_{s=1}^\infty   dx^\un
A_\un{}^{n_1\ldots n_{s-1}} (y|x) p_{n_1}\ldots p_{n_{s-1}}\,, \ee
\be
\BB(p,y|x)=\sum^\infty_{s=0} \BB{}^{n_1\ldots n_s} (y|x) p_{n_1}\ldots
p_{n_s}\,.
\ee
A massless spin $s$ field is described
by the 1-forms $A(p,y|x)$ and 0-forms $\BB(p,y|x)$
which are, respectively, of order $s-1$ and $s$ in $p$.

The equations (\ref{d2})  are
invariant under the gauge transformations
\be \label{gof}
\delta A = D \epsilon =d \epsilon + \hp^{-1}[A , \epsilon ]_*\,, \qquad
\delta \BB = \hp^{-1} [\BB , \epsilon ]_* \,,
\ee
where $\epsilon (p,y|x)$ is an arbitrary function of its
arguments. As will be shown
in section \ref{Linearization},
some of these
gauge parameters are responsible for gauging away redundant
degrees of freedom while the rest become the usual
lower spin and higher spin gauge symmetry
parameters (like, e.g., standard Yang-Mills symmetry).

A global HS symmetry of the vacuum solution (\ref{A0}) and
(\ref{b0}) has  parameters $\epsilon (p,y)$
that commute to $\half p^2$. As shown in
section \ref{Linearization}, this means that they are described by the polynomials
$$\epsilon (p,y)=
\epsilon^{n_1 \ldots n_{s-1}\,,m_1 \ldots m_t}
 p_{n_1}\ldots p_{n_{s-1}}
y_{m_1}\ldots y_{m_t}\,,
$$
 where the parameters
$
\epsilon^{n_1 \ldots n_{s-1}\,,m_1 \ldots m_t}\,
$
have the symmetry properties of Young tableaux
\begin{picture}(60,20)(0,-5)
{\linethickness{.250mm}
\put(00,00){\line(1,0){60}}%
\put(00,05){\line(1,0){60}}%
\put(00,-5){\line(0,1){10}}%
\put(05,-5){\line(0,1){10}} \put(10,-5){\line(0,1){10}}
\put(15,-5){\line(0,1){10}} \put(20,-5){\line(0,1){10}}
\put(25,-5){\line(0,1){10}} \put(30,-5){\line(0,1){10}}
\put(35,-5){\line(0,1){10}} \put(40,00.0){\line(0,1){05}}
\put(45,00.0){\line(0,1){05}} \put(50,00.0){\line(0,1){05}}
\put(55,00.0){\line(0,1){05}} \put(60,00.0){\line(0,1){05}}
\put(00,-5){\line(1,0){35}}%
}
\put(21,6.2){\scriptsize  ${s-1}$}
\put(21,-10){\tiny  ${t}$}
\end{picture}\,\,\,
with $s-1$ cells in the
upper row and
$t$ cells in the second one. This set of symmetries precisely matches
that of the HS algebra of \cite{non}. The two HS algebras are not
isomorphic, however. The algebra considered in this paper is
a contraction of the (off-shell)
$AdS$ HS algebra of \cite{non} pretty much like
Poincar\`{e} algebra is a contraction of the $AdS$ algebra.

The system (\ref{d2}) is obviously consistent
and nonlinear. Despite it has the trivial form of
zero-curvature and covariant constancy
conditions\footnote{Let us note that the form of the
system (\ref{d2}) is analogous to that of the basis
system of the Fedosov quantization prescription \cite{fed}.
The difference is that we use its restriction to the
Lagrangian base subsurface $M^d$ of the corresponding
noncommutative $2d$-dimensional phase space. Analogous
system was also considered in \cite{BNW} in a different context.},
it describes nontrivial models
 including gravity and Yang-Mills.
This is possible because the 0-form $\BB$ has a nonzero
vacuum expectation value $\BB_0$ (\ref{b0}) so that
the gauge symmetries are spontaneously broken and the
flat 1-form connection $A(p,y|x)$ cannot be completely
gauged away by the leftover unbroken gauge symmetries.
To prove that the system (\ref{d2})  describes
appropriately off-shell relativistic fields
it is enough to check that
it is doing so at the linearized level. This will be shown
in section \ref{Linearization}.

The off-shell HS system (\ref{d2}) admits simple truncations
to the lower spin sectors of spins zero, one and two.
This may sound surprising because it  is
usually claimed that lower-spin fields
are sources to the HS fields (see e.g. \cite{Gol}).
Analogously to the argument on the (ir)relevance of the
$AdS$ background, although
this is indeed true on-shell, it may not be so
off-shell where no equations of motion are imposed
on the dynamical fields and, therefore, no obstruction results from
a nontrivial $\sigma_-$ cohomology associated with
the equations of motion
to impose restrictions on spin spectra and/or geometry of the
background space-time. In other words, as long as
there are no nontrivial dynamical equations
on dynamical  fields, there is also no question on how their right hand
sides (currents) are built of other fields. In the absence of dynamical
field equations, i.e. when all relations are some constraints, whatever is
added to the right hand sides of the constraints
it can always be absorbed into a
redefinition of auxiliary fields.

In the case of gravity, the connection 1-form
$A(p,y|x)$ is linear in $p^n$
\be
A(p,y|x)= A^n (y|x) p_n \,.
\ee
Because every commutator takes away at least one
power of $p^n$, the left hand side of the zero curvature
equation in (\ref{d2}) is also linear in $p^n$.
With the help of the star product (\ref{weyl}) one obtains
\be
\label{da2}
d A^n (y|x) +
A^m (y|x)\wedge \f{\partial }{\partial y^m}A^n (y|x)=0\,.
\ee
This is the zero curvature equation for
the Lie algebra of vector fields in the variables $y^n$.

The spin two 0-form $\BB(p,y|x)$ is bilinear
in $p^n$. Let us also introduce a spin zero
(i.e., $p^n$-independent) ``dilaton" field by setting
\be
\BB(p,y|x)= \BB^{mn} (y|x) p_m p_n + \BB (y|x)\,.
\ee
The covariant constancy condition on $\BB$ in (\ref{d2})
amounts to
\bee
\label{db2}
d \BB^{mn} (y|x) &+&
A^l (y|x)\f{\partial }{\partial y^l} \BB^{mn} (y|x)\nn\\
&-&\f{\partial }{\partial y^l} A^m (y|x)\BB^{ln} (y|x)-
\f{\partial }{\partial y^l} A^n (y|x)\BB^{lm} (y|x)=0\,
\eee
and
\be
\label{db0}
d \BB (y|x) +
A^l (y|x)\f{\partial }{\partial y^l} \BB (y|x)
+\f{\hp^2}{4}
\f{\partial^2 }{\partial y^l \partial y^n} A^m (y|x)
\f{\partial }{\partial y^m }\BB^{ln} (y|x)=0\,.
\ee
The  equation (\ref{db2}) is the covariant constancy condition
for the second rank symmetric tensor $ \BB^{mn} (y|x)$.
The  equation (\ref{db0}) contains two types of terms.
The $\hp$--independent terms represent the
vector field covariant constancy condition in
the scalar representation. The $\hp$--dependent term
describes a source term built of the gravitational
fields. Since the system remains consistent for any value
of $\hp$, one can consistently set $\hp$ equal to zero.
This limit corresponds to the Poisson bracket realization of the
algebra of vector fields, where the equation (\ref{db0}) just
describes a scalar field in the external gravitational field.

That the last term in the equation (\ref{db0}) can be added
is very interesting, however. It describes a nontrivial co-cycle of
the algebra of vector fields with polynomial coefficients.
Having three derivatives, it is somewhat reminiscent of the central
extension of the Witt algebra to Virasoro.

The system is invariant under the $y-$vector field transformations
(``diffeomorphisms" with the coefficients polynomial in $y$)
gauged in $x$ space-time
\be
\label{dea2}
\delta A^n (y|x)=d \epsilon^n (y|x) +
A^m (y|x)\f{\partial }{\partial y^m}\epsilon^n (y|x)\,
-\epsilon^m (y|x)\f{\partial }{\partial y^m} A^n (y|x)\,,
\ee
\be
\label{deb2}
\delta \BB^{mn} (y|x)=
\f{\partial }{\partial y^l} \epsilon^m (y|x)\BB^{ln} (y|x)+
\f{\partial }{\partial y^l} \epsilon^n (y|x)\BB^{lm} (y|x)
-\epsilon^l (y|x)\f{\partial }{\partial y^l} \BB^{mn} (y|x)
\,,
\ee
\be
\label{deb0}
\delta \BB (y|x) =
-\epsilon^l (y|x)\f{\partial }{\partial y^l} \BB (y|x)
-\f{\hp^2}{4}
\f{\partial^2 }{\partial y^l \partial y^n} \epsilon^m (y|x)
\f{\partial }{\partial y^m }\BB^{ln} (y|x)=0\,,
\ee
where $\epsilon^l (y|x)$ is an arbitrary function of $x^\un$
and $y^n$ that expands in power series of $y^n$.

Let us note that if $A(p,y|x)$ and  $\BB(p,y|x)$ were
matrices (i.e. took values in some noncommutative associative algebra),
the property that every commutator takes away at least one
power of $p^n$ would not be true any more\footnote{Indeed, a wedge product of
matrix valued connection 1-form $A^i{}_j (p)$ linear in $p$ will
generically contain terms bilinear in $p^n$ because
$
[a\otimes x \,,b\otimes y ]=
\half( [a,b] \otimes\{x \,,y\} +\{a\,,b\}\otimes [x,y])\,,
$
for $a,b\in A_1$ and $x,y \in A_2$ where $A_{1,2}$ are two associative
algebras, i.e. when the matrix commutator is not zero, the star product
anticommutator of the functions of  $y$ and $p$ appears.}.
This means that if colored spin two particles appear, this is
only possible in the presence of HS fields, which conclusion is
in agreement with the results of \cite{CW}.

As explained in sections \ref{Linearization} and
\ref{Nonlinear off-mass-shell unfolding},
the equations (\ref{da2}) and (\ref{db2}) provide unfolded
formulation of off-shell nonlinear gravity. The equation (\ref{db0})
with $\hp \neq 0$ extends it in a non-trivial way
to a system with dilaton.  It can be further extended to a system
of fields with spins $s\leq 1$.

A spin one massless field is described by the 1-form
\be
{\cal A} = \hp^{-1} dx^\un {\cal A}_\un(y|x)
\ee
 along with the 0-form $p_n \BB^n (y|x)$,
both  taking values in a
matrix algebra of a non-Abelian Yang-Mills theory.
Spin zero matter fields are described by a 0-form $ \BB(y|x)$
in some representation of the Yang-Mills algebra. The corresponding
equations read as
\be
\label{da1}
d {\cal A} (y|x) + A^m (y|x)\wedge \f{\partial }{\partial y^m}{\cal A}(y|x)
+{\cal A}(y|x)\wedge {\cal A}(y|x)=0\,,
\ee
\bee
\label{db1}
d \BB^{m} (y|x)  -\f{\p}{\p y_m } {\cal A}(y|x) &+&
A^l (y|x)\f{\partial }{\partial y^l} \BB^{m} (y|x)
-\f{\partial }{\partial y^n} A^m (y|x)\BB^{n} (y|x)\nn\\
&+&[{\cal A}(y|x)\,,\BB^{m} (y|x) ]=0\,
\eee
and
\be
\label{db1m}
d \BB (y|x) +
A^l (y|x)\f{\partial }{\partial y^l} \BB (y|x)
+{\cal A}(y|x)(\BB (y|x))=0\,,
\ee
where $A^l (y|x)$ describes the background gravitational field,
$[,]$ is the usual commutator in the matrix
algebra where the Yang-Mills fields ${\cal A}(y|x)$ and $\BB^n(y|x)$
take their values while ${\cal A}(y|x)(\BB (y|x))$ is a result
of the action of the
Yang-Mills field ${\cal A}(y|x)$ on the scalar field $\BB(y|x)$
according to the representation of the Yang-Mills algebra carried
by $\BB(y|x)$.

The corresponding gauge transformation law is
\be
\label{dea1}
\delta{\cal A} (y|x)=
d \epsilon (y|x) +A^m \f{\partial }{\partial y^m} \epsilon (y|x)
- \epsilon^m (y|x) \f{\partial }{\partial y^m}{\cal A} (y|x)
+[{\cal A}(y|x)\,, \epsilon (y|x)]\,,
\ee
\bee
\label{deb1}
\delta \BB^{m} (y|x) =\f{\p}{\p y_m } {\epsilon}(y|x)&-&
\epsilon^l (y|x)\f{\partial }{\partial y^l} \BB^{m} (y|x)
+\f{\partial }{\partial y^n} \epsilon^m (y|x)\BB^{n} (y|x)\nn\\
&+&
[\BB^{m} (y|x)\,,\epsilon(y|x) ]=0\,,
\eee
and
\be
\label{deb1m}
\delta \BB (y|x) =-
\epsilon^l (y|x)\f{\partial }{\partial y^l} \BB (y|x)
-\epsilon(y|x)(\BB (y|x))=0\,.
\ee
It forms a semidirect product of the Yang-Mills
algebra with $y$--dependent parameters $\epsilon (y|x)$
(forming an ideal) and
the algebra of $y$--space vector fields $\epsilon^m (y|x)$
polynomial in $y$.

Let us note that the unfolded formulation of Yang-Mills theory
and gravity presented here is deeply related to earlier
important work on a reformulation of these fundamental theories.
In particular, relevance
of the $y$-dependent symmetries to the description of Yang-Mills
theory was originally found in the sigma-model approach
of Ivanov and Ogievetsky
\cite{ivog}. The idea that twistor transform of
the full nonlinear Yang-Mills equations can be achieved via doubling
of space-time coordinates in the Yang-Mills connection
was put forward by Witten \cite{WYM} (see also \cite{YIG}).
Ivanov observed  \cite{ivanov} that Cartan forms  built of
the goldstonion fields of the sigma-model approach to  Yang-Mills
theory satisfy the zero-curvature
equation (\ref{da1}). Analogous construction for the case of
gravity was developed by Pashnev in \cite{pashnev}, where it
was shown that the relevant infinite-dimensional algebra is the
algebra of $y$-space diffeomorphisms while  the Cartan 1-forms,
built of the goldstonion fields of the sigma-model approach to gravity,
satisfy the zero-curvature equation (\ref{da2}).
Note that, from the perspective of our approach,
the goldstonion fields of \cite{ivog,ivanov,pashnev} appear
in the exponential parametrization of the gauge function $g(y|x)$
of the pure gauge representation  $A=g^{-1} d g$
for  solutions of the zero curvature equations
(\ref{da1}) or (\ref{da2}).

As will be explained elsewhere, it is also possible to
include spin 1/2 matter fields into our lower-spin model.
On the other hand, if any (boson or fermion)
HS field with $s>2$ is added, the
system (\ref{d2})  requires an infinite set
of HS fields to be included. This is so because the commutator of
the connection $A(p,y|x)$ of order $s_1-1$ in $p$
with the 0-form $\BB(p,y|x)$ of order $s_2$ in $p$
contributes to the part of order $s_1 +s_2 -2$
in $p$ in the equation (\ref{d2}) for $\BB$.
This means that spin $s_1$ and
spin $s_2$ fields produce a source for spin $s= s_1 +s_2 -2$. As a result,
 any spin greater than two gives rise to an infinite
tower of higher and higher spins. In other words if a HS field
is present in the system under consideration, no truncation to a
subsystem with a finite number of spins is available.
It is possible, however, to truncate the system to only
even spins where every even spin appears just once.

The property that HS fields
form infinite towers is a consequence of the structure
of the star product algebra. On the other hand,
since unfolding of only
Bianchi identities with no dynamical equations imposed is always
possible, it should be  possible to
formulate an off-shell HS system for any given spin in the background
gravitational and/or Yang-Mills fields.
As discussed in more detail in section \ref{contr}, this is achieved
by replacing the star product algebra with the commutative algebra of
functions of commuting variables $y^n$ and $p_n$
in the spin $s\neq 2$ sector and with the Poisson algebra
in the spin two sector of connections linear in $p$.

\section{Free Fields Unfolded}
\label{Free Fields Unfolded}

To explain how the equations (\ref{d2}) describe
a nonlinear HS system we recall in this section
the unfolded formulation of free totally symmetric
massless fields of any  spin elaborated in
\cite{Fort1} for the $d=4$ case
and then extended  to any dimension in \cite{LV,5d}.
We first consider important examples of spin one
in subsection \ref{Free massless spin one}
and spin two in subsection \ref{unfgrav}, coming to the
general case in subsection \ref{Free any spin}.

\subsection{Free spin one}
\label{Free massless spin one}
A spin one gauge field is
described by a vector potential $a_\un (x)$.
One writes
\be
\label{maxten}
da=e_0^n\wedge e_0^m C_{n,m}\,,
\ee
where $e_0^n  = dx^n$ is the frame 1-form of the background
Minkowski space and
the antisymmetric tensor $C_{n,m}=-C_{m,n}$ parametrizes
components of the field strength. It is depicted by the Young tableau
\begin{picture}(09,13)
{\linethickness{.250mm}
\put(05,05){\line(1,0){05}}%
\put(05,10){\line(1,0){05}}%
\put(05,0){\line(1,0){05}}%
\put(05,0){\line(0,1){10}}%
\put(10,0.0){\line(0,1){10}}
}
\end{picture}\,. The first derivative of the field
strength \be \label{dmax} d C_{n,m}= e_{0}^p C_{n,m;p}
\ee
has the structure
\be
\begin{picture}(09,13)
{\linethickness{.250mm}
\put(05,05){\line(1,0){05}}%
\put(05,10){\line(1,0){05}}%
\put(05,0){\line(1,0){05}}%
\put(05,0){\line(0,1){10}}%
\put(10,0.0){\line(0,1){10}}
}
\end{picture}\,\,\,
\otimes
\begin{picture}(5,20)(0,-2)
{\linethickness{.250mm}
\put(00,00){\line(1,0){5}}%
\put(00,5){\line(1,0){5}}%
\put(00,0){\line(0,1){5}}%
\put(05,0){\line(0,1){5}}%
}
\end{picture}   \,\,\,=\,\,\,
\begin{picture}(09,18)(0,2)
{\linethickness{.250mm}
\put(05,05){\line(1,0){05}}%
\put(05,10){\line(1,0){05}}%
\put(05,15){\line(1,0){05}}%
\put(05,0){\line(1,0){05}}%
\put(05,0){\line(0,1){15}}%
\put(10,0.0){\line(0,1){15}}
}
\end{picture}\,\,\,\oplus
\begin{picture}(09,13)
{\linethickness{.250mm}
\put(05,05){\line(1,0){10}}%
\put(05,10){\line(1,0){10}}%
\put(05,0){\line(1,0){05}}%
\put(05,0){\line(0,1){10}}%
\put(10,0.0){\line(0,1){10}}
\put(15,05){\line(0,1){05}}
}
\end{picture}\,\,\,\quad.
\ee
The Bianchi identities for (\ref{maxten}) imply however that the
part that contains antisymmetrization over three indices in
(\ref{dmax}) must be zero:\,\,
\begin{picture}(09,18)(0,2)
{\linethickness{.250mm}
\put(05,05){\line(1,0){05}}%
\put(05,10){\line(1,0){05}}%
\put(05,15){\line(1,0){05}}%
\put(05,0){\line(1,0){05}}%
\put(05,0){\line(0,1){15}}%
\put(10,0.0){\line(0,1){15}}
}
\end{picture}\,\,\,=0. This means that \be \label{011} d
C_{n,m}= e_{0}^p (C_{np,m}-C_{mp,n})\,,
\ee
where $C_{np,m}$ has
symmetry properties of the hook Young tableau being
otherwise arbitrary, i.e. $C_{mn,p}=C_{nm,p}$  and
symmetrization over
three indices gives zero: $C_{nm,p}+C_{np,m}+C_{mp,n} =0$. The
equation (\ref{011}) is the first step of the
unfolding of spin one
dynamics in the 0-form sector. The next step is to analyse Bianchi
identities for (\ref{011}) that impose restrictions on the first
derivative of $C_{np,m}$. The process continues indefinitely leading
to the following chain of differential relations \be \label{unf01}
dC_{m_1\ldots m_{l}, n }=e_0^p\left( (l+1) C_{m_1\ldots m_{l}p, n} +
C_{m_1\ldots m_{l}n, p}\right) \ee with the 0-forms $C_{m_1\ldots
m_{l}, n }$ described by the  Young tableau
\,\,\,
\begin{picture}(45,15)(0,-5)
{\linethickness{.250mm}
\put(00,00){\line(1,0){45}}%
\put(00,05){\line(1,0){45}}%
\put(00,-5){\line(0,1){10}}%
\put(05,-5){\line(0,1){10}}
\put(10,0){\line(0,1){5}}
\put(15,0){\line(0,1){5}}
\put(20,00.0){\line(0,1){05}}
\put(25,00.0){\line(0,1){05}}
\put(30,00.0){\line(0,1){05}}
\put(35,00.0){\line(0,1){05}}
\put(40,00.0){\line(0,1){05}}
\put(45,00.0){\line(0,1){05}}
\put(00,-5){\line(1,0){5}}%
}
\put(16,6.2){\small  ${l}$}
\end{picture}\,\,\,\, ,
{\it i.e.,} $C_{m_1\ldots m_{l}, n}$ is symmetric in the indices $m$ and
such that symmetrization over all $l+1$ indices gives zero \be
\label{1sym} C_{\{m_1\ldots m_{l}, m_{l+1}\}}=0\,. \ee

The system (\ref{maxten}) and (\ref{unf01}) gives off-shell
unfolded formulation of spin one field because in its derivation we
did not make use of the Maxwell equations. The latter require the
first derivative of the Maxwell tensor to be traceless. In other
words, Maxwell equations require the tensor $C_{mn,k}$ on the right
hand side of (\ref{011}) to be traceless. {}From (\ref{unf01}) it
follows then that all tensors $C_{m_1\ldots m_{l}, n}$ are also
traceless \be \label{1tr} C_{m_1\ldots m_{l}, n}\eta^{m_1 m_2}
=0\,,\qquad C_{m_1\ldots m_{l}, n}\eta^{m_1 n} =0\,. \ee Therefore,
the system (\ref{maxten}) and (\ref{unf01}) along with the
conditions (\ref{1sym}), (\ref{1tr}) gives unfolded formulation of
the Maxwell equations, i.e., the on-shell unfolded formulation
of a spin one free massless particle.

The traceful (traceless)  0-forms
$ C_{m_1\ldots m_{l}, n}$ form a basis of the
space of all gauge invariant derivatives in the
off-shell (on-shell) Maxwell theory. They can be
described by a generating function
\be
\label{C1}
C(p,y|x)= \sum_{l=1}^\infty  C_{m_1\ldots m_{l}, n}
y^{m_1}\ldots y^{m_l} \,p^n \,
\ee
satisfying the Young condition
\be
\label{you}
y^n\frac{\partial}{\partial p^n}C(p,y|x)=0\,
\ee
equivalent to (\ref{1sym}).
To put the free spin one system on-shell is equivalent to imposing
the tracelessness conditions
\be
\label{trs1}
\frac{\partial^2}{\partial y^m \partial y_m} C(p,y|x)=0\,,\qquad
\frac{\partial^2}{\partial y^m \partial p_m} C(p,y|x)=0\,.
\ee

Suppose now that we want to reformulate analogously a non-Abelian
Yang-Mills theory. To this end one replaces the left hand side of
(\ref{maxten}) by the non-Abelian Yang-Mills field strength
\be
\label{ymten} da+\half [a\,,a] =e_0^n\wedge e_0^m C_{n,m}\,,
\ee
where both $a$ and $C_{n,m}$ take values in a Yang-Mills Lie
algebra. The Bianchi identities then give a relation analogous to
(\ref{011})
\be D C_{n,m}= e_{0}^p (C_{np,m}-C_{mp,n})\,,
\ee
where $D$ is the Yang-Mills covariant derivative in the adjoint
representation. The
analysis of the Bianchi identity for this equation is more
complicated because $D^2$ is itself proportional to $C_{n,m}$ by
virtue of (\ref{ymten}). As a result, the equations (\ref{unf01})
with higher $l$ must acquire nonlinear corrections. To the best of
our knowledge a full nonlinear form of the unfolded spin one system
was not known so far both in the off-shell and in the
on-shell cases. One of the results of this paper consists
of the observation that the equations (\ref{d2}) solve
the off-shell Yang-Mills problem when restricted to the spin one case.
We will also show in section \ref{field equations for lower spins}
how the nonlinear Yang-Mills equations
look like in these terms.

\subsection{Linearized gravity}
\label{unfgrav}

The set of fields in Einstein-Cartan's formulation of gravity
consists of the frame field $e_\um{}^m$ and the Lorentz connection
$\o_\m{}^{nm}$. One supposes that the torsion constraint (\ref{tor})
is satisfied in order to express the Lorentz connection in terms of the
frame field. The Lorentz curvature can be expressed as
$R^{mn}=e_k\wedge e_l\,R^{mn\,;\,kl}$, where $R^{mn\,;\,kl}$ is a rank four
tensor antisymmetric
in the pair of indices $mn$ and $kl$, having the symmetries
of the tensor product\,
\begin{picture}(09,13)
{\linethickness{.250mm}
\put(05,05){\line(1,0){05}}%
\put(05,10){\line(1,0){05}}%
\put(05,0){\line(1,0){05}}%
\put(05,0){\line(0,1){10}}%
\put(10,0.0){\line(0,1){10}}
}
\end{picture}\,\,
$\otimes$\,\,
\begin{picture}(09,13)
{\linethickness{.250mm}
\put(05,05){\line(1,0){05}}%
\put(05,10){\line(1,0){05}}%
\put(05,0){\line(1,0){05}}%
\put(05,0){\line(0,1){10}}%
\put(10,0.0){\line(0,1){10}}
}
\end{picture}\,.
 The
algebraic Bianchi identity $e_n\wedge  R^{mn}=0$, which follows from the
zero torsion constraint, implies that the tensor $R^{mn\,;\,kl}$
possesses the symmetries of the Riemann tensor, {\it i.e.}
$R^{[mn\,;\,k]l}=0$. This means that it carries an irreducible
representation of $GL(d)$ characterized by the Young tableau
\begin{picture}(13,12)(0,0)
{\linethickness{.250mm}
\put(00,10){\line(1,0){10}}
\put(00,05){\line(1,0){10}}
\put(00,00){\line(1,0){10}}
\put(00,00){\line(0,1){10}}
\put(05,00){\line(0,1){10}}
\put(10,00){\line(0,1){10}}
}
\end{picture}\,\,.

For our purpose, it is more convenient to use the symmetric
basis. In this convention, one supplements the zero torsion
condition  (\ref{tor}) with the equation \be
R^{mn}\,=\,e_k\,\wedge e_l\,C^{mk,\,nl}\,,
\label{Einst}
\ee
where the $0$-form $C^{mk,\,nl}$ is symmetric in the pairs $mk$ and
$nl$ and satisfies the algebraic relation
$C^{\{mk,\,n\}l}=0\,,$ which implies the window Young symmetry
\begin{picture}(13,12)(0,0)
{\linethickness{.250mm}
\put(00,10){\line(1,0){10}}
\put(00,05){\line(1,0){10}}
\put(00,00){\line(1,0){10}}
\put(00,00){\line(0,1){10}}
\put(05,00){\line(0,1){10}}
\put(10,00){\line(0,1){10}}
}
\end{picture}\,\,
in the symmetric basis.
To start the unfolding of linearized gravity around the
Minkowski background one linearizes the equation of
(\ref{Einst}) to
\be
\label{HSpin2}
R_1^{mn}\,=\,e_{0\;k}\,\wedge e_{0\;l}\,C^{mk,\,nl} \,.
\ee
To unfold this equation one has to add the equations
containing the differential of the Riemann $0$-form $C^{mk,\,nl}$.
Since we do not want to impose any additional dynamical restrictions
on the system, the only restrictions on the derivatives of the Riemann
$0$-form $C^{mk,\,nl}$ result from the Bianchi identities
applied to (\ref{HSpin2}).

{\it A priori}, the first Lorentz covariant derivative of the
Riemann tensor is a rank five tensor in the following representation
\begin{eqnarray}
\begin{picture}(13,12)(0,0)
{\linethickness{.250mm}
\put(00,10){\line(1,0){10}}
\put(00,05){\line(1,0){10}}
\put(00,00){\line(1,0){10}}
\put(00,00){\line(0,1){10}}
\put(05,00){\line(0,1){10}}
\put(10,00){\line(0,1){10}}
}
\end{picture}
\otimes
\begin{picture}(5,20)(0,-2)
{\linethickness{.250mm}
\put(00,00){\line(1,0){5}}%
\put(00,5){\line(1,0){5}}%
\put(00,0){\line(0,1){5}}%
\put(05,0){\line(0,1){5}}%
}
\end{picture}   \,\,\,=\,\,\,
\begin{picture}(13,12)(0,0)
{\linethickness{.250mm}
\put(00,10){\line(1,0){10}}
\put(00,05){\line(1,0){10}}
\put(00,00){\line(1,0){10}}
\put(00,-5){\line(1,0){05}}
\put(00,-5){\line(0,1){15}}
\put(05,-5){\line(0,1){15}}
\put(10,00){\line(0,1){10}}
}
\end{picture}
\oplus
\begin{picture}(13,12)(0,0)
{\linethickness{.250mm}
\put(15,5){\line(0,1){05}}
\put(00,10){\line(1,0){15}}
\put(00,05){\line(1,0){15}}
\put(00,00){\line(1,0){10}}
\put(00,00){\line(0,1){10}}
\put(05,00){\line(0,1){10}}
\put(10,00){\line(0,1){10}}
}
\end{picture}\,\,\,\,
\label{Yougdec}\\\nonumber\end{eqnarray}
of $gl(d)$. The Bianchi identity applied
to (\ref{HSpin2}) implies that the part of the derivative of the
Riemann tensor antisymmetrized in three indices must vanish. As a
result, in the decomposition (\ref{Yougdec}) of the Lorentz
covariant derivative of the Riemann tensor, the first term vanishes
and the second term is arbitrary. This is equivalent to say that \be
dC^{mk,\,nl}\,=\,{e_0}_f\,(2C^{mkf,\,\,nl}+C^{mkn,\,\,lf}+C^{mkl,\,\,nf})\,,
\label{unfgrav2} \ee
 where the right hand side is fixed
 by the Young symmetry properties of the left hand side
modulo an overall normalization coefficient. This equation is the
first step of the unfolding procedure in the sector of 0-forms.
$C^{mkf,\,\,nl}$ is irreducible under $gl(d)$. The
analysis of the Bianchi identities goes on  indefinitely
resulting in the infinite set of equations
\be
dC^{m_1\ldots m_{k+2},\,n_1n_2}= e_{0\;l}\,\Big(\,(k+2)\,C^{m_1
\ldots m_{k+2}l,\, n_1 n_2}\,+\,C^{m_1 \ldots m_{k+2} n_1,\,n_2
l}\,+
\,C^{m_1 \ldots m_{k+2} n_2,\,n_1 l}\,\Big)
\label{unfoldingggravity}
\ee
$(0\leq k<\infty)$, where the fields $C^{m_1\ldots m_{k+2},\,n_1n_2}$ are
described by the two-row
Young tableau {\it
i.e.}
\be
\label{s2yu}
C^{\{m_1\ldots m_{k+2},\,n_1\}n_2}=0\,.
\ee
 As expected, the system (\ref{unfoldingggravity}) is
 consistent with $d^2C^{m_1\ldots m_{k+2},\,n_1n_2}=0$.

Analogously to the spin zero and spin one  cases,
the meaning of the 0-forms
$C^{m_1\ldots m_{k+2},\,n_1n_2}$
is that they form a basis in the
space of all gauge invariant combinations of the derivatives of the
spin two gauge field. The set of 0-forms
$C^{m_1\ldots m_{k+2},\,n_1n_2}$  can be conveniently
described by a generating function
\be
C(p,y|x)= \sum_{l=2}^\infty  C_{m_1\ldots m_{l}, n_1 n_2}
y^{m_1}\ldots y^{m_l} \,p^{n_1}p^{n_2} \,,
\ee
satisfying the Young condition that has
the same form (\ref{you}) as in the spin one case.

The vacuum Einstein equations which put the system on-shell
state that $C_{mn,kl}$ is traceless.
In other words, the Riemann tensor is equal on-shell to the
Weyl tensor. By virtue of Bianchi identities, at the linearized level,
all derivatives of the Weyl tensor are also traceless on-shell. (For
more detail see, e.g., \cite{solv}). In other words, unfolded
formulation of the linearized spin two equations is given by the
equations (\ref{HSpin2}) and (\ref{unfoldingggravity}) at the
condition that the fields $C^{ m_1\ldots m_{k+2},\,n_1n_2}$ satisfy
(\ref{s2yu}) and are all traceless, $ \eta_{m_1m_2}C^{m_1m_2\ldots
m_{k+2},\,n_1n_2}=0\,, $ which condition has the same form (\ref{trs1})
as in the spin one case. The traceless  0-forms $C^{m_1\ldots
m_{k+2},\,n_1n_2}$ form a basis in the space of all on-mass-shell
nontrivial gauge invariant combinations of the derivatives of the
spin two gauge field.

To extend the free field unfolded formulation of the massless spin
two field to the nonlinear level
one has to replace $d$ and $e_0^n$ with the Lorentz
covariant derivative $D^L$ and the dynamical frame field $e^n$,
respectively.
Since $D^LD^L$ is the Riemann tensor, Bianchi identities for the
covariantized equations (\ref{unfoldingggravity}) will require
nonlinear corrections to these equations analogous to those in the
Yang-Mills theory. The nonlinear corrections to the covariantized
field equations (\ref{unfoldingggravity}) quadratic in the 0-forms
$C$ were found in \cite{tri} for the d=4 case. In this paper we
find  both off-shell and
on-shell nonlinear formulation for spin two in any dimension.

\subsection{Free symmetric massless fields of any spin}
\label{Free any spin}

Free unfolded HS field equations can be formulated
 \cite{Fort1,LV,5d} in terms of the
1-form gauge fields which generalize those of the
Cartan formulation
of gravity. Namely, for a massless spin $s$, one introduces a set of
1-forms $dx^\m\o_{\m}^{ ~n_1 \ldots n_{s-1}, \, m_1 \ldots m_t}$ which
have the symmetry of the two-row Young tableaux
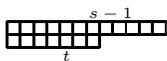
\begin{picture}(60,20)(0,-5)
{\linethickness{.250mm}
\put(00,00){\line(1,0){60}}%
\put(00,05){\line(1,0){60}}%
\put(00,-5){\line(0,1){10}}%
\put(05,-5){\line(0,1){10}} \put(10,-5){\line(0,1){10}}
\put(15,-5){\line(0,1){10}} \put(20,-5){\line(0,1){10}}
\put(25,-5){\line(0,1){10}} \put(30,-5){\line(0,1){10}}
\put(35,-5){\line(0,1){10}} \put(40,00.0){\line(0,1){05}}
\put(45,00.0){\line(0,1){05}} \put(50,00.0){\line(0,1){05}}
\put(55,00.0){\line(0,1){05}} \put(60,00.0){\line(0,1){05}}
\put(00,-5){\line(1,0){35}}%
}
\put(31,6.2){\tiny  ${s-1}$}
\put(21,-10){\tiny  ${t}$}
\end{picture}\,\,
with $s-1$ cells in the first row  and any $0\leq t \leq s-1$  cells
in the second row
with respect to the fiber indices $n$ and $m$.

In the on-shell system considered in
\cite{Fort1,LV,5d}, the connections \hfil\\
$dx^\m \o_{\m}^{ ~n_1 \ldots n_{s-1}, \, m_1 \ldots m_t}$ are traceless in
the fiber indices $n$ and $m$. As shown in \cite{SSS,solv},
the off-shell version is obtained by relaxing the tracelessness
condition. We first discuss the off-shell case with traceful
connections.

The frame-like HS field is that with $t=0$
\be
\label{fr}
dx^\um e_ {\um}^{n_1 \ldots n_{s-1}}=
dx^\um \o_{\um}^{n_1 \ldots n_{s-1}}\,.
\ee
The Lorentz connection-like auxiliary field has $t=1$, i.e.
$\o_{\m}^{ ~n_1 \ldots n_{s-1}, \, m}$. The fields with $t>1$ are
called extra fields. They appear for higher spins with $s>2$. The
set of gauge fields $\o_{\m}^{ ~n_1 \ldots n_{s-1}, \, m_1 \ldots
m_t}$ generalizes the sets of spin one and spin two gauge fields
to any spin, containing them as particular cases. The flat space
linearized HS curvatures have the form
\be R_1^{n_1 \ldots n_{s-1},\, m_1 \ldots
m_{t}}\,=\,D^L_0\o^{n_1 \ldots n_{s-1},\, m_1 \ldots
m_{t}}\,+\,e_{0\;k}\, \o^{n_1 \ldots n_{s-1},\, m_1 \ldots
m_{t}k}\,, \label{tauminus}\ee
where $D^L_0$ and $e_0^n$ are,
respectively, the Lorentz covariant differential and the frame
1-form of the background Minkowski space ($D^L=d$ and $e^n_0=dx^n$ in
Cartesian coordinates). These linearized curvatures generalize to
any spin the spin one Maxwell field strength and spin two linearized
torsion ($t=0$) and Riemann tensor ($t=1$).

The linearized HS gauge transformations have the form
\be
\delta \o^{n_1 \ldots n_{s-1},\, m_1 \ldots
m_{t}}\,=\,D^L_0
\epsilon^{n_1 \ldots n_{s-1},\, m_1 \ldots m_{t}}
\,+\,e_{0\;k}\, \epsilon^{n_1 \ldots n_{s-1},\, m_1 \ldots
m_{t}k}\,, \label{transs}
\ee
where the 0-form
$\epsilon^{n_1 \ldots n_{s-1},\, m_1 \ldots m_{t}}(x)$ is an
arbitrary gauge parameter with the symmetry properties of the
two-row Young tableau with $s-1$ cells in the first row
and $t$ cells in the second row.

The HS analogue of the equation (\ref{maxten}) for spin one and
the equation (\ref{HSpin2}) for spin two is
 \be
R_1^{n_1 \ldots n_{s-1},\, m_1 \ldots
m_{t}}\,=\,\d_{t,\,s-1}\,\,\,e_{0\;k}\wedge \,e_{0\;l}\,\,
C^{n_1 \ldots
n_{s-1}k,\, m_1 \ldots m_{s-1}l}\,\qquad (0\leq t\leq s-1)\,.
\label{onmshell}
\ee
The $0$-form $C^{n_1 \ldots n_s,\, m_1 \ldots m_s}$ is the spin-$s$
Riemann-like tensor. It is
characterized by a rectangular two-row Young tableau\,
\begin{picture}(60,20)(0,-5)
{\linethickness{.250mm}
\put(00,00){\line(1,0){60}}%
\put(00,05){\line(1,0){60}}%
\put(00,-5){\line(0,1){10}}%
\put(05,-5){\line(0,1){10}} \put(10,-5){\line(0,1){10}}
\put(15,-5){\line(0,1){10}} \put(20,-5){\line(0,1){10}}
\put(25,-5){\line(0,1){10}} \put(30,-5){\line(0,1){10}}
\put(35,-5){\line(0,1){10}} \put(40,-5.0){\line(0,1){10}}
\put(45,-5.0){\line(0,1){10}} \put(50,-5.0){\line(0,1){10}}
\put(55,-5.0){\line(0,1){10}} \put(60,-5.0){\line(0,1){10}}
\put(00,-5){\line(1,0){60}}%
}
\put(31,6.2){\scriptsize  ${s}$}
\end{picture}\,.
The key property of this system is \cite{solv} that it imposes
no restrictions on the frame-like connection (\ref{fr}).
All connections
$\o_{\m}^{ ~n_1 \ldots n_{s-1}, \, m_1 \ldots m_t}$ with $t>0$
are expressed
via its order-$t$ derivatives modulo Lorentz-like Stueckelberg
gauge symmetries (\ref{transs}) with
$\epsilon^{ ~n_1 \ldots n_{s-1}, \, m_1 \ldots m_t}$ with $t>0$.
The Riemann tensor $C^{n_1 \ldots n_{s},\, m_1 \ldots m_{s}}$
represents order-$s$ gauge invariant combinations of the derivatives
of the spin $s$ frame-like field and coincides with the analogous
tensor found in the metric-like formalism by de Wit and Freedman
\cite{WF}.

The analysis of the Bianchi identities of (\ref{onmshell}) works for
any spin $s\geq 2$ in a way analogous to gravity. The final result
is the following equation which has a form of a covariant constancy
condition
\bqn 0=\widetilde{D}_0C^{n_1 \ldots n_{s+k},\, m_1 \ldots
m_s}&\equiv& D_0^LC^{n_1 \ldots n_{s+k},\, m_1 \ldots
m_s}\nn\\&&\ls\ls\ls\ls\ls\ls\ls\ls
-\,e_{0\;k}\Big((2+k)C^{n_1 \ldots n_{s+k}k,\, m_1 \ldots
m_s}\,+\,s\,
C^{n_1 \ldots n_{s+k} \{m_1,\,m_2 \ldots m_s\}k}\Big)
\label{unfHS}\eqn
$0\leq k< \infty$,
where $C^{n_1 \ldots n_{s+k},\, m_1 \ldots m_s}$ are $gl(d)$
 tensors  characterized by the Young tableaux
\begin{picture}(60,20)(0,-5)
{\linethickness{.250mm}
\put(00,00){\line(1,0){60}}%
\put(00,05){\line(1,0){60}}%
\put(00,-5){\line(0,1){10}}%
\put(05,-5){\line(0,1){10}} \put(10,-5){\line(0,1){10}}
\put(15,-5){\line(0,1){10}} \put(20,-5){\line(0,1){10}}
\put(25,-5){\line(0,1){10}} \put(30,-5){\line(0,1){10}}
\put(35,-5){\line(0,1){10}} \put(40,00.0){\line(0,1){05}}
\put(45,00.0){\line(0,1){05}} \put(50,00.0){\line(0,1){05}}
\put(55,00.0){\line(0,1){05}} \put(60,00.0){\line(0,1){05}}
\put(00,-5){\line(1,0){35}}%
}
\put(31,6.2){\tiny  ${s+k}$}
\put(21,-10){\tiny  ${s}$}
\end{picture}\,\,.
They describe off-shell nontrivial $k$-th derivatives of the
spin-$s$ Riemann-like tensor, thus forming a basis in the space of
gauge invariant combinations of derivatives of a spin $s$ HS gauge
field. The system (\ref{unfHS}) is an extension of the spin
zero, spin one and spin two off-shell systems considered above to
any spin. Let us stress that, for $s \geq 1$, the infinite
system of equations (\ref{unfHS}) describes constraints that
expresses higher tensors
$C^{n_1 \ldots n_{s+k},\,m_1 \ldots m_s}$ in terms
of derivatives of the Riemann-like tensors
$C^{n_1 \ldots n_{s} ,\,m_1 \ldots m_s}$ along with
consequences  of
(\ref{onmshell}) by the Bianchi identities.
Note that (\ref{onmshell}) makes no sense
for $s=0$  because there is no spin 0 gauge potential while
(\ref{unfHS}) with $s=0$ reproduces the unfolded
spin zero off-shell equation (\ref{un0}).
The fact that the system
(\ref{onmshell}) and (\ref{unfHS}) imposes no differential
conditions on the frame-like connection (\ref{fr})
and the spin zero 0-form $C$ \cite{solv}
will be referred to as Central off-mass-shell theorem.

The set of HS connection 1-forms
$\o^{n_1 \ldots n_{s-1},\, m_1 \ldots m_{t}}$ and
Riemann 0-forms
$C^{n_1 \ldots n_{s+k},\, m_1 \ldots m_{s}}$
can be described by the generating functions
\be
\o (p,y|x)= \sum_{l,k}
\o_{m_1\ldots m_{l}, n_1 \ldots n_{k}}
p^{m_1}\ldots p^{m_l} \,y^{n_1}\ldots y^{n_k} \,
\ee
and
\be
C (p,y|x)= \sum_{l,k}  C_{m_1\ldots m_{l}, n_1 \ldots n_{k}}
y^{m_1}\ldots y^{m_l} \,p^{n_1}\ldots p^{n_k} \,,
\ee
which satisfy the conditions
\be
\label{you1}
\tau_+ \o(p,y|x)=0\,,
\ee
\be
\label{you0}
\tau_- C(p,y|x)=0\,,
\ee
where the operators
\be
\label{sl2}
\tau_+ = p^n \f{\p}{\p y^n} \,,\qquad
\tau_- = y^n \f{\p}{\p p^n} \,,\qquad
\tau_0 =[\tau_- ,\tau_+ ]=
y^n \f{\p}{\p y^n}-p^n \f{\p}{\p p^n}
\ee
form a $sp(2)$ algebra. The conditions (\ref{you1}) and (\ref{you0})
imply that symmetrization over any $l+1$ indices
of the coefficients $\o_{m_1\ldots m_{l}, n_1 \ldots n_{k}}$ and
$\C_{m_1\ldots m_{l}, n_1 \ldots n_{k}}$ gives zero. This is
equivalent to the fact that they are described by the
two-row Young tableaux with $l$ cells in the first row and $k$
cells in the second row.

For a fixed spin $s$ one has
\be
\label{sp1}
p^n\frac{\partial}{\partial p^n}\o(p,y|x)=(s-1) \o(p,y|x)\,,
\ee
\be
\label{sp0}
p^n\frac{\partial}{\partial p^n}C(p,y|x)=s C(p,y|x)\,.
\ee

Following  \cite{Fort1,LV,5d},
to put the system on-shell one requires
both the HS connections $\o(p,y|x)$ and 0-forms
$C(p,y|x)$ to be traceless, that is harmonic in the fiber variables
$y$ and $p$
\be
\label{trs2}
\frac{\partial^2}{\partial y^m \partial y_m} \o(p,y|x)=0\,,\qquad
\frac{\partial^2}{\partial y^m\partial p_m} \o(p,y|x)=0\,,\qquad
\frac{\partial^2}{\partial p^m \partial p_m} \o(p,y|x)=0\,.
\ee
\be
\label{trs0}
\frac{\partial^2}{\partial y^m\partial y_m} C(p,y|x)=0\,,\qquad
\frac{\partial^2}{\partial y^m \partial p_m} C(p,y|x)=0\,,\qquad
\frac{\partial^2}{\partial p^m \partial p_m} C(p,y|x)=0\,.
\ee
The resulting field equations (\ref{onmshell}) generalize
(\ref{HSpin2}) of linearized gravity.
That they indeed describe free equations of
motion for any spin equivalent to the equations,
which follow from the action principle of Fronsdal \cite{Fr},
is the content of the so called Central on-mass-shell theorem
proved originally in \cite{Fort1,unf} for the $4d$
case and then in \cite{LV,5d} for any dimension (see also \cite{solv}
for more detail).

Note that the metric-like field of the Fronsdal formalism arises as the
completely symmetric part of the frame field
 $ \varphi_{\m_1 \ldots \m_s}=e_{\{\m_1 ,\, \m_2\ldots
\m_s\}}\,,$ where all fiber indices have been lowered using the
frame field $e_0{}^m_\m$.
From the  fiber index tracelessness of the on-shell
HS frame field it follows automatically that the field $\varphi_{\m_1
\ldots \m_s}$ is double traceless as required in the original
Fronsdal formulation \cite{Fr} (see \cite{sor} for pedagogical reviews
of the Fronsdal formulation).

\section{$sp(2)$ generators}
\label{$sp(2)$ generators}

It is well known that the star product (\ref{weyl})
describes the associative product of the Weyl
ordered (i.e., totally symmetrized) polynomials of oscillators.
The following formulae are simple
consequences of the star product law (\ref{weyl})
 \be\label{star1}
 [p_n, f(p,y)]_* =\hp
\f{\p}{\p y^n} f(p,y)\,,
\ee
\be
\label{star2}
 \{p_n, f(p,y)\}_* = 2    p_n f(p,y)\,.
\ee

Using these relations along with  the identity
$[a^2\,, b]_* \equiv \{a,[a,b]_*\}_*$  one observes
that
\be
\label{p2}
[p^2 , f ]_* = 2\hp p^n \f{\p}{\p y^n} f =2 \hp \tau_+ (f)\,.
\ee
More generally, the $sp(2)$ generators  (\ref{sl2})
admit the following realization in terms of
the star product algebra
\be
\tau_\pm (f)= \f{1}{\hp} [t_\pm \,, f]_*\,,\qquad
\tau_0 (f) = \f{1}{\hp} [t_0 \,, f]_*
\ee
with
\be
t_-= - \half y_n y^n\,,\qquad
t_+=  \half p_n p^n\,,\qquad
t_0=   y_n p^n\,.
\ee
{}From the associativity of the star product
it  follows in particular that, given
two lowest vectors $f$ and $g$
satisfying $[t_-\,, f]_* = [t_- \,,g]_* =0$,
their star product is also a lowest vector:
$[t_- \,,f*g]_* =[t_- \,,f]_* *g +f*[t_- \,,g]_* =0$.

In the subsequent analysis we will use the following simple
lemmas:

\vspace{0.2cm}
\noindent
{\it Lemma 1}

\noindent
Any polynomial $X(p,y)$ can be uniquely
decomposed as
\be
\label{decom}
X(p,y)=X(p,y)\Big|_{V_-} + \tau_+ \Big( X (p,y)\Big|_-\Big)
\,, \qquad \tau_-
\Big (X(p,y)\Big|_{V_-}\Big )=0\,.
\ee

\noindent
{\it Proof:}
Since the $sp(2)$ generators $\tau_\pm$
do not change a degree of a polynomial of $p$ and $y$,
the space of functions
$X (p,y)$ decomposes into an infinite direct sum of
finite-dimensional $sp(2)$  submodules spanned by
homogeneous polynomials of different degrees. Elements of
these submodules can
be generated by $\tau_+$ from lowest weight vectors
annihilated by $\tau_-$, i.e.,
any $X (p,y)$ can be represented as
\be
\label{hw}
X (p,y) = \sum_{q\geq 0} (\tau_+ )^q X_q (p,y) \,,\qquad
\tau_- X_q(p,y) =0\,.
\ee
Then $\xv= X_0(p,y)$ and $\xm =
\sum_{q\geq 1} (\tau_+ )^{q-1} X_q (p,y)$. $\Box$

Note that although $ \tau_+ \Big (\xm \Big )$ in (\ref{decom})
is defined uniquely, $ \xm$ is defined modulo elements in
$Ker \,\tau_+$. In what follows we will also use notation
$X\Big |_{V_+}$ and $X\Big |_{+}$ for the decomposition
analogous to (\ref{decom}) with the roles of $\tau_-$ and $\tau_+$
exchanged.

\vspace{0.2cm}
\noindent
{\it Lemma 2}

\noindent
Let
\be
\label{l2cond}
\tau_+ (X(p,y))=0\,,\qquad \tau_0 (X(p,y))=\alpha X(p,y)\,,
\ee
where $\alpha$ is a number. Then
\be
\label{x}
\xv = X\,, \qquad if\quad \alpha = 0\,,
\ee
\be
\xv = 0\,, \qquad if\quad \alpha \neq 0\,.
\ee
This lemma follows from the fact that,
for  finite-dimensional $sp(2)$ modules,
a lowest vector can the same time be a highest vector
only for a trivial module.

\newpage
\noindent
{\it Lemma 3}

\noindent
Let
\be
\label{t0+}
\tau_0 X^+ (p,y) = Y^+(p,y)\,,
\ee
with $t^{-1}\,Y^+(t^{-1}p,t y)$ being a polynomial of $t$ or
\be
\label{t0-}
\tau_0 X^- (p,y) = Y^-(p,y)\,,
\ee
with $t^{-1}\,Y^-(tp,t^{-1} y)$ being a  polynomial of $t$. Then
\be
\label{lem+}
X^+ (p,y) = \int_0^1 dt t^{-1}  Y^+(t^{-1}p,t y)
\ee
and
\be
\label{lem-}
X^- (p,y) =- \int_0^1 dt t^{-1}  Y^-(t p,t^{-1} y)\,
\ee
are particular solutions of the equations (\ref{t0+}) and
(\ref{t0-}), respectively. We skip the proof as it is elementary.

\section{Linearization}
\label{Linearization}

In this section we show that the nonlinear HS equations
(\ref{d2}) reproduce correctly the off-shell HS system
at the linearized level.

Let us analyze the equations (\ref{d2})  and the
gauge transformations (\ref{gof}) perturbatively.
Taking into
account (\ref{A0}) and (\ref{b0}) and choosing Cartesian coordinates
with $D^L =d$ and $e^n = dx^n$ one obtains to the first order
\be
\label{da11} (d+ dx^n \f{\p}{\p y^n}) A_1 (p,y|x) =0\,,
\ee
\be
\label{db11}
\tau_+ A_1 (p,y|x)=
(d+ dx^n \f{\p}{\p y^n})\BB_1(p,y|x)\,,
\ee
where we made use of (\ref{p2}).

The linearized  gauge transformations are
\be
\label{dga1}
\delta_0 A_1 (p,y|x) =(d+ dx^n \f{\p}{\p y^n})\epsilon (p,y|x)\,,
\ee
\be\label{gb1}
 \delta_0 \BB_1 (p,y|x) =\tau_+ \epsilon (p,y|x)\,.
 \ee

Let us now show that the equations (\ref{da11}) and (\ref{db11}) along
with the gauge transformations (\ref{dga1}) and
(\ref{gb1}) reproduce Central
off-mass-shell theorem of subsection \ref{Free any spin}.
 To illustrate the
idea we start with the lower spin examples of spin zero and
spin  one.

\subsection{Spin zero and spin one}
\label{Spin zero and spin one}

For spin zero there is no gauge field $A$ and gauge symmetry
parameter $\epsilon$. The spin zero part of the form $\BB$ is
$p^n$--independent. The only nontrivial equation is therefore
(\ref{db11}) at $p^n =0$, i.e. $(d+ dx^n \f{\p}{\p y^n})\BB_1
(0,y|x)=0$, which is just the unfolded equation (\ref{xu2}) of
subsection \ref{Free massless scalar}.

In the spin one case we have $p$-independent gauge potential
$dx^\un A_\un (y|x)$ and 0-form $\BB_1$ linear in $p^n$, i.e
\be \BB_1
=\sum_{k=0}^\infty  p^n y^{m_1}\ldots y^{m_k} \BB_{1\ m_1\ldots
m_k\,;\,n }\,.
\ee
The gauge transformation (\ref{gb1}) with a
$p$--independent gauge parameter has the form \be \delta \BB_{1\
m_1\ldots m_k\,;n }= \epsilon_{m_1\ldots m_{k}n}\,,
 \ee
where  $\epsilon_{m_1\ldots m_{l}}(x)$ is an arbitrary
totally symmetric tensor. This means that the totally symmetric
part in $\BB_{1\ m_1\ldots
m_k\,;n }$ is pure gauge. Gauge fixing
the totally symmetric part of $\BB_{1\ m_1\ldots m_k\,;n }$
to zero is equivalent to say that
\be \label{bc} \BB_{1\
m_1\ldots m_k\,;n}= C_{m_1\ldots m_{l}, n }\,,
\ee
where the field
$C_{m_1\ldots m_{l}, n }$
has the symmetry properties of
the Young tableau\,\,\,
\begin{picture}(45,20)(0,-5)
{\linethickness{.250mm}
\put(00,00){\line(1,0){45}}%
\put(00,05){\line(1,0){45}}%
\put(00,-5){\line(0,1){10}}%
\put(05,-5){\line(0,1){10}}
\put(10,0){\line(0,1){5}}
\put(15,0){\line(0,1){5}}
\put(20,00.0){\line(0,1){05}}
\put(25,00.0){\line(0,1){05}}
\put(30,00.0){\line(0,1){05}}
\put(35,00.0){\line(0,1){05}}
\put(40,00.0){\line(0,1){05}}
\put(45,00.0){\line(0,1){05}}
\put(00,-5){\line(1,0){5}}%
}
\put(16,6.2){\small  ${l}$}
\end{picture}\,.
These fields form the set of 0-forms which describe all
off-shell gauge invariant derivatives of the spin one field
as discussed in subsection \ref{Free massless spin one}.
Thus, in this gauge
\be
\BB_1(p,y|x)=C(p,y|x)\,,\qquad \tau_- (C) =0\,,
\label{c1}
\ee
where $C(p,y|x)$ is the spin one 0-form (\ref{C1})
satisfying the condition (\ref{you}).

The leftover gauge symmetry is described by the $y$-independent
gauge parameter $\epsilon (x)$ which does not contribute to the
variation of the spin one  0-form $\BB_1$. $\epsilon (x)$ is the
usual spin one gauge parameter.

Let us now apply the operator $\tau_-$ to the both sides
of (\ref{db11}), taking into account (\ref{c1}) and
that $A(p,y|x)=A(y|x)$ is $p$--independent for spin 1.
This gives
\be \label{ab} y^n \f{\p}{\p
y^n}A_1 (y|x)=- dx^n \f{\p}{\p p^n}C (p,y|x)\,,
\ee
which means by {\it Lemma 3} that
\be
\label{ab1}
A_1 (y|x)= a(x) - dx^n \f{\p}{\p p^n} \int_0^1
dt t^{-1}C (p,t y|x)\,, \ee
where $a(x)$ is an independent 1-form
to be identified with the electromagnetic potential. Note that there
is no pole in $t$ on the right hand side of the equation (\ref{ab1})
because a field $C (p, y|x)$ satisfying (\ref{you0}) is at
least linear in $y^n$ (since  a Young tableau with one cell in the
second row associated with $p^m$ must contain at least one cell in
the first row associated with $y^n$). Plugging this back into the
equation (\ref{db11}) one obtains
\be
\label{unf0} (d+ dx^n \f{\p}{\p
y^n})C (p,y|x) + dx^n p^m \f{\p^2}{\p p^n \p y^m} \int_0^1 dt
t^{-1}C (p,t y|x)=0\,.
\ee
By an appropriate rescaling
of the coefficients of the expansion of $C(p,y|x)$
in powers of $y$ and $p$
this reproduces the equation (\ref{unf01}) for $s=1$.

Finally, it remains to plug (\ref{ab1}) into (\ref{da11}).
One observes that, applying
the operator $\tau_+$  to the both sides of
(\ref{da11}) gives identity once the equation (\ref{db11}) is true.
Therefore, a nontrivial part of (\ref{da1}) is
contained in its $y^n$ independent sector. Setting $y=0$ in
(\ref{da11}) one obtains
\be
\label{ch11}
da(x) = dx^m \wedge dx^n \f{\p^2}{\p p^n \p y^m} C(p,y)\Big |_{p=y=0}\,,
\ee
which is equivalent to the equation (\ref{maxten}).
The equations (\ref{unf0}) and ({\ref{ch11})
form the free off-shell spin one unfolded system.

\subsection{Any spin}

Let us now consider the general case.
Taking into account that
\be
\label{gtrlin}
\delta_0 \BB_1 (p,y|x) = \tau_+ \epsilon (p,y|x)
\ee
and using {\it Lemma 1} we see that it is possible to
gauge away all components of $\BB_1 (p,y)$ except for
those of the form
\be
\label{vacb}
\BB_1 (p,y|x) = C(p,y|x)\,,\qquad \tau_- C (p,y|x) =0\,,
\ee
i.e., $C(p,y|x)=\BB_1 (p,y|x)\Big |_{V_-}$. The field $C(p,y|x)$ is
just  the generating function for the HS 0-forms
satisfying the Young (anti)symmetry condition (\ref{you0}).

{}From (\ref{gtrlin}) it also follows that the remaining gauge symmetry
parameters satisfy the highest weight condition
\be
\label{tauple}
\tau_+ \epsilon (p,y|x)=0\,,
\ee
{\it i.e.,} the coefficients of the expansion
\be
\epsilon (p,y) = \sum_{p\geq q} \epsilon_{n_1\ldots n_p\,, m_1\ldots m_q}
p^{n_1} \ldots p^{n_p} y^{m_1} \ldots y^{m_q}\,
\ee
are described by two-row Young tableaux for which $p$ is
associated with the first row and $y$ is associated with the second
row. These are the gauge parameters
of the frame-like formulation           for symmetric HS
fields \cite{LV} summarized in subsection
\ref{Free any spin}.

Let us now consider equation (\ref{db11}). It reconstructs
$A_1 (p,y|x)$ in terms of $C (p,y|x)$ modulo solutions of
the homogeneous equation, that is
\be
A_1 (p,y|x) =\o (p,y|x) +\tilde{A}(p,y|x)\,,
\ee
where $\o (p,y|x)$ is an arbitrary $sp(2)$ highest weight
polynomial satisfying (\ref{you1}) to be identified with
the HS gauge potentials,
while $\tilde{A}$ is a particular solution of the equation (\ref{db11}).
It is convenient to look for $\tilde{A}$ among the lowest weight
vectors
\be
\label{tila}
\tau_-
\tilde{A} (p,y|x)=0\,.
\ee
Applying the operator
$
\tau_-
$
to the both sides of the equation (\ref{db11}) one obtains
\be
\tau_0(\tilde{A}) =-
 dx^n \f{\p}{\p p^n}C(p,y|x)\,.
\ee
To solve this equation we apply {\it Lemma 3}
(\ref{t0+}), (\ref{lem+}), taking into account that,
for $C(p, y|x)$ satisfying (\ref{vacb}),
a power of $p$ cannot be higher than that of $y$.
This  reconstructs
$\tilde{A}$ in the form compatible with (\ref{tila})
\be
\label{abs}
A(p,y|x)=\o (p,y|x) -
dx^n \int_0^1 dt
\f{\p}{\p p^n}C(t^{-1}p,t y|x)\,.
\ee

Plugging this expression back into (\ref{db11}) one obtains
\be
\label{dbc}
(d+ dx^n \f{\p}{\p y^n})C(p,y|x)+dx^n
p^m \f{\p}{\p y^m}
\int_0^1 dt \f{\p}{\p p^n}C(t^{-1}p,t y|x)=0\,.
\ee
By an appropriate rescaling
of the coefficients of the expansion of $C(p,y|x)$
in powers of $y$ and $p$ this reproduces the equation
(\ref{unfHS}) for the chains of 0-forms of any spin.

Finally, it remains to plug (\ref{abs}) into (\ref{da11}).
Like in the spin one case, the application of
the operator $\tau_+$  to the both sides of
(\ref{da11}) gives identity once (\ref{db11}) is true.
This means that a nontrivial part of (\ref{da11}) is
contained in the kernel of $\tau_+$. As a result,
it is easy to see that the nontrivial part of the
equation (\ref{da11}) is
\be
\label{dac}
d\o(p,y|x) +dx^n\f{\p }{\p y^n}\o(p,y|x)
= dx^m\wedge dx^n \f{\p^2}{\p p^n \p y^m} C^l (p,y|x) \,,
\ee
where $C^l (p,y|x)$ is the lowest tensor
part of $C(p,y|x)$ that is annihilated both by
$\tau_-$ and by $\tau_+$  thus being a $sp(2)$ singlet
\be
\tau_- C^l(p,y|x)=0\,,\qquad \tau_+ C^l(p,y|x) =0\,,\qquad
\tau_0 C^l(p,y|x)=0\,,
\ee
which means that it is described by a rectangular Young tableau,
{\it i.e.}
 $C^l(p,y|x)$ is the Maxwell tensor for $s=1$, Riemann tensor for $s=2$
and their HS analogues for higher spins. (Note that
$C^l(t^{-1}p,ty|x)=C^l(p,y|x)$.) The equation (\ref{dac})
reproduces the chain of constraints (\ref{onmshell}). Thus, the
equations (\ref{dac}), (\ref{dbc}) form the unfolded off-shell system
for a free spin $s$ massless field provided that $\o(p,y|x)$ and
$C(p,y|x)$ are, respectively, of degrees $s-1$ and $s$ in $p^n$.

Let us note that the analysis of
this section is algebraically
analogous to the analysis of free HS fields performed in \cite{BGST}
within the BRST approach, where $sp(2)$ generators (constraints)
also play the key role. In particular, the fact that the
on-shell version of the equations (\ref{da11}) and (\ref{db11})
describes properly linearized HS dynamics was shown by the authors
of \cite{BGST}.
The important difference is that the analysis of \cite{BGST}
uses the standard BRST language of ``states" where the BRST operator
acts, that requires a doubled number of oscillators compared to
our approach where ``states" are replaced by the associative star
product algebra. Among other things, the advantages of our
formalism are that all formulae remain valid in any coordinate system
in Minkowski space by replacing de Rham differential $d$ with the
Lorentz covariant derivative $D^L$ and $dx^n$ by the frame 1-form $e^n$
and, most important, that it admits
a natural generalization to the interacting case.

\section{Nonlinear off-shell analysis}
\label{Nonlinear off-mass-shell unfolding}

To reproduce unfolded HS constraints it is important
in our approach that
$ A (p,y|x)$ and $\BB(p,y|x)$ contain the zero-order parts
(\ref{A0}) and (\ref{b0}). In the nonlinear analysis we will not
single out the vacuum part of the  gravitational field, setting
\be
\label{gr1}
 A(p,y|x) = dx^\un \left (
e_{\un}{}^n (x)p_n +\go_{\un}{}^{nm}(x)p_n y_m
+   {A}_{\un\,1}(p,y|x)\right )
\,,
\ee
\be
\label{gr0}
 \BB(p,y|x) = t_+  +  \BB_1(p,y|x)
\,.
\ee
Here
$e_{\un}{}^n (x)p_n$, $ \go_{\un}{}^{nm}(x)p_n y_m$
and of course the flat Minkowski metric $\eta^{mn}$ in $t_+$
are not supposed to be small\footnote{Of course, this requirement is
needed when discussing a phase with a nondegenerate metric tensor
as in the space we used to live in. One can in principle think of
different phases where what we call gravitational metric field is
degenerate. This interesting option is in fact very much in spirit
of the unfolded dynamics and may be related to models with invisible
extra dimensions as discussed in \cite{Mar}.}.
The fields $ {A}_{\un\,1}(p,y|x)$ and
$ {\BB}_{1}(p,y|x)$ are treated as fluctuations.
$ {A}_{\un\,1}(p,y|x)$ does not
include the gravitational part associated with the fields $e^n$ and
$\o^{mn}$ in (\ref{gr1}), i.e we require
\be
\label{grav}
 {A}_{\un\,1}(p,y|x)\Big|_{V_+}=0\,
\qquad \mbox{if} \qquad
\Big ( p^n\f{\partial}{\partial p^n}-1\Big ){A}_{\un\,1}(p,y|x)=0\,.
\ee

The gauge transformation law is
\be
\delta A(p,y|x) = D^L \epsilon (p,y|x)  +
e^n\f{\p}{\p y^n}\epsilon (p,y|x) + [A_1 (p,y|x)\,,\epsilon(p,y|x)]_*\,,
\ee
\be
\label{db111}
\delta \BB_1(p,y|x) =
\tau_+ (\epsilon (p,y|x)) + [\BB_1 (p,y|x)\,,\epsilon(p,y|x)]_*\,,
\ee
where $D^L$ is the Lorentz covariant derivative
\be
D^L= d +\o^{mn}(y_n\f{\p}{\p y^m} + p_n\f{\p}{\p p^m} )\,.
\ee

Since the gauge transformation law (\ref{db111})
has the form (\ref{gtrlin}) in the leading order,
it is still possible to impose the gauge condition (\ref{vacb}),
which we write in the form
\be
\label{t-c}
\BB_1 (p,y|x) = C(p,y|x)\,,\qquad [ t_- \,, C]_* =0\,.
\ee

The field equations (\ref{d2}) now read as
\be
\label{DA}
R^{mn}p_m y_n +R^m p_m +D^L A_1(p,y|x) +e^n \f{\p}{\p y^n} A_1(p,y|x)
+(A_1*\wedge A_1)(p,y|x) =0\,,
\ee
\be
\label{DB}
D^L C(p,y|x) +e^n \f{\p}{\p y^n} C(p,y|x) -\tau_+ (A_1(p,y|x)) +
[A_1(p,y|x)\,,C(p,y|x)]_*=0\,,
\ee
where $ R^m $ and $R^{mn}$ are, respectively,
the  torsion tensor (\ref{tor}) and the Riemann tensor (\ref{riem}).

Since the  equation (\ref{DA}) is the compatibility condition
for (\ref{DB}), a significant part of
the information contained in (\ref{DA}) is a consequence of that
contained in (\ref{DB}). Let us show that a part of
the  equation (\ref{DA}) independent of (\ref{DB})
belongs to $Ker \,\tau_+$. The proof is by induction.

Actually, suppose that the content of the equation (\ref{DA})
has been checked up to an order $q$ in powers of
$\BB_1=C$, while that of
the equation (\ref{DB}) has been checked up to the order $q+1$.
(One can choose $q=-1$ as the starting point in which case the
equation (\ref{DB}) implies that $A_1(p,y|x)$ belongs to
$Ker \,\tau_+$.)
Consider the order $q+1$ part of the equation
\be
\label{BD2}
[\BB\,,(d+A)\wedge(d+A)]_* =0\,.
\ee
Since this is a consequence of the equation
\be
d\BB +[A,\BB]_* =0
\ee
which by assumption is true up to an order $q+1$ in $C$,
it follows that the equation (\ref{BD2}) is also
satisfied up to the same order. Taking into account
(\ref{gr1}), where $\BB_1(p,y|x)$ is of order one, and that
the equation $(d+A)\wedge(d+A)=0$ was by assumption analysed
 up to order $q$ it follows from (\ref{BD2}) that
\be
\tau_+ ((d+A)\wedge * (d+A)) =0
\ee
is true up to order $q+1$ as a consequence of the other equations.
$\Box$

Thus, the only nontrivial part of the equation (\ref{DA})
that remains to be analysed is
that in $Ker \,\tau_+$. It
can be written in the form
\be
\label{DAP}
R^{mn}p_m y_n     +R^m p_m +
\Big (D^L A_1(p,y|x) + e^n \f{\p}{\p y^n} A_1(p,y|x) +
(A_1*\wedge A_1)(p,y|x)\Big )\Big |_{V_+}
 =0\,.
\ee

So, the equations (\ref{DB}) and (\ref{DAP})
are equivalent to the equations
(\ref{d2}) with the gauge condition (\ref{t-c}).
 Let us note that the general analysis remains true
if all fields take values in a matrix algebra ({\it i.e.,} carry
matrix indices) provided that the gravitational field lies in its
center  ({\it i.e.,} is proportional to the unit element of this
matrix algebra).

Before considering the general case, let us discuss
the important lower spin examples of gravity, Yang-Mills theory
and a scalar field.

\subsection{$s\leq 2$}
\label{sleq2}

What simplifies the analysis of the
lower spins $s\leq 2$ is  that the property
(\ref{tila}) of the linearized analysis remains
true in all orders. Indeed, taking into account
(\ref{t-c}), the algebraic part of the
equation (\ref{DB}), that does not contain $DC(p,y|x)$,
results from the application of $\tau_-$ to the both sides
of (\ref{DB}). This gives
\bee
\label{eqc}
\tau_0 ( A_1 (p,y|x))&=&-e^n \f{\p}{\p p^n} C(p,y|x)
-\tau_+ (\tau_- ( A_1 (p,y|x)))\nn\\
&+&
[\tau_- (A_1(p,y|x))\,,C(p,y|x)]_*=0\,.
\eee
One can now iterate this equation to reconstruct  $A_1 (p,y|x)$
in terms of $ C(p,y|x)$.  According to
the original equation (\ref{DB}), it should reconstruct
$A_1 (p,y|x)$ in terms of $ C(p,y|x)$ up to
zero modes of $\tau_+$ identified with the dynamical
gauge fields $\o(p,y|x)$. The appearance of the latter
fields in the last term on the right hand side of (\ref{eqc})
does not allow us to find a solution with
$A_1 (p,y|x)$ being a lowest weight element for the general spin.
For spin two, however, the zero mode
has been already separated into $e^n$ and $\o^{nm}$ so that,
by the condition (\ref{grav}), it is not contained in
$A_1 (p,y|x)$. For spin one,
the zero mode $a(x)$ is independent of $p^n$ and $y^n$, i.e. it itself
satisfies the lowest weight condition in a trivial way.

As a result, for spins $s\leq 2$ we solve (\ref{eqc}) the same way as
in the linear problem to obtain
\be
\label{lac}
A_1(p,y|x)=a (x) - e^n \int_0^1 dt
\f{\p}{\p p^n}C(t^{-1}p,t y|x)\,,
\ee
where $a(x)$ is the Yang-Mills gauge field. The last two terms
in (\ref{eqc}) drop out because
\be
\tau_- (A_1(p,y|x) )=0\,.
\ee

Let us stress that the analysis of this section
works for the non-Abelian
case where $a(x)$ and all spin one
0-forms (i.e., those linear in $p^n$) take values in
a Yang-Mills algebra $g$, while the spin zero 0-forms independent
of $p$ are in some representation of $g$.

To analyse the content of the equation (\ref{DAP})
 we observe that a $sp(2)$ singlet part $a(x)$ of the solution
(\ref{lac}) is the Yang-Mills field. The $C$--dependent
part in (\ref{lac}) belongs to $Ker\, \tau_-$ but contains
no singlets because its degree in
$y$ is strictly higher than that of $p$ as a result
of differentiation over $p$ in (\ref{lac}), that means
that it contains only  eigenvectors  of $\tau_0$ with
strictly positive eigenvalues. Taking into account
{\it Lemma 2} along with the properties that $(i)$ star product of two
elements $a_{1,2}$ satisfying the lowest weight conditions
 $[t_- \,, a_{1,2}]_*=0,$ is also lowest weight and $(ii)$
$[t_0\,, a_1 *a_2 ]_* = (\alpha_1 +\alpha_2) a_1 *a_2$ if
$[t_0\,, a_i]_* = \alpha_i  a_i $ (i=1,2), we observe that
 all $C$--dependent terms in $A_1$ do not contribute to (\ref{DAP})
 except for the term $e^n \f{\p}{\p y^n} A_1(p,y|x)$ that decreases
 the $y$ degree by one unit. The terms with the Yang-Mills connection
 $a(x)$ form the Yang-Mills field strength
 \be
 R(x)= da(x) +a(x)\wedge a(x)\,.
\ee

As a result, the equation (\ref{DAP}) simplifies to
\be
\label{curls}
R^{mn}(x) p_m  y_n +R^m(x) p_m + R(x) =
 e^n\wedge e^m  \f{\p^2}{\p y^n \p p^m} C^l(p,y|x)\,,
\ee
where $C^l(p,y|x)$ is the $sp(2)$ singlet part of the 0-form
$C(p,y|x)$ described by rectangular Young tableaux, so that
for the spins two, one and zero
\be
C^l(p,y|x)= C^{kl,nm}(x) y_k y_l p_n p_m+
C^{n,m}(x) y_n p_m + C(x)\,.
\ee
The spin two and spin one parts contribute to the equation
(\ref{curls}). The meaning of this equation is now obvious.
It implies that torsion tensor is zero, thereby expressing the
Lorentz connection via the frame field $e^n$, and identifies
$C^{nm,kl}(x)$ and  $ C^{n,m}(x)$ with the components of the
Riemann and Yang-Mills curvatures imposing no conditions on the
latter.

Now we plug the expression (\ref{lac}) into the equation (\ref{DB}).
The terms with the Yang-Mills connection form the full
Lorentz-Yang-Mills derivative
\be
D^{LYM}= D^L+[a, \ldots ]\,.
\ee
The equation (\ref{DB}) now takes the form
\bee
\label{DBL}
D^{LYM}C(p,y|x) &+&e^n\int_0^1 dt\Big ( \f{\p}{\p y^n} C(p,y|x) +
\tau_+ \Big (
\f{\p}{\p p^n}C(t^{-1}p,t y|x)\Big )\nn\\&-&
[\f{\p}{\p p^n}C(t^{-1}p,t y|x))\,,C(p,y|x)]_*\Big )=0\,.
\eee
This equation provides a fully consistent nonlinear deformation of the
linearized equation (\ref{unfHS}). Together with
(\ref{curls}) it gives the unfolded form of nonlinear
off-shell constraints
for the fields of spins two, one and zero.

Let us mention a subtle point. In the formulation
with the star product we use, there is a contribution of
the terms built of the spin two Riemann tensor and
its derivatives to the sector of spin zero 0-forms
in the equation (\ref{DBL}), which is
\be
\hp^2 e^n \f{\p^4}{\p p^n\p p^m\p y^k \p y^l}
C(p, y|x)\f{\p^3}{\p y_m\p p_k \p p_l }C(p,y|x) \Big |_{p=0}\,.
\ee
Although this term disappears in the ``semiclassical" limit
$\hp \to 0$ and can therefore be consistently neglected off-shell,
its appearance indicates that HS fields form sources for lower
spin fields which effect looks to be inevitable in the on-shell
theory in $AdS_d$ where it is not possible to take a semiclassical
limit of the star product algebra.

\subsection{Any spin}

Using the decomposition (\ref{decom}) we obtain from (\ref{DB})
two equations.
One is the differential equation on $C(p,y|x)$
\be
D^L C(p,y|x) +e^n \Big (\f{\p}{\p y^n} C(p,y|x)\Big )\Big |_{V_-} +
([A_1(p,y|x)\,,C(p,y|x)]_*)\Big |_{V_-}=0\,.
\ee
Another one
\be
\tau_+ (A_1(p,y|x)) =
\tau_+ \Big (\Big (e^n \f{\p}{\p y^n} C(p,y|x)+
[A_1(p,y|x)\,,C(p,y|x)]_* \Big )\Big |_-\Big )\,
\ee
is the algebraic equation that reconstructs the gauge
connection $A_1(p,y|x)$ in terms of $C(p,y|x)$ and HS gauge
connections $\o(p,y|x)$ satisfying (\ref{you1}).

Since  $\tau_- (C)=0$, the following identity is true
\be
\tau_- \Big (e^n \f{\p}{\p y^n} C(p,y|x)
+\tau_+ \Big ((\tau_0 )^{-1} e^n \f{\p}{\p p^n} C(p,y|x)
\Big )\Big )=0\,,
\ee
which means that
\be
(e^n \f{\p}{\p y^n} C(p,y|x))\Big |_{V_-} =
e^n \f{\p}{\p y^n} C(p,y|x))
+\tau_+ ((\tau_0 )^{-1} e^n \f{\p}{\p p^n} C(p,y|x))\,,
\ee
\be
\tau_+ (e^n \f{\p}{\p y^n} C(p,y|x) )\Big |_- =-
\tau_+\Big ( (\tau_0 )^{-1} e^n \f{\p}{\p p^n} C(p,y|x)\Big )\,.
\ee
By virtue of these identities and using {\it Lemma 3} one obtains
\bee
\label{DC}
D^L C(p,y|x)&+&
e^n \f{\p}{\p y^n} C(p,y|x))
+\tau_+\Big (e^n \int_0^1 dt
 [\f{\p}{\p p^n}C(t^{-1}p,t y|x)\Big )\nn\\&+&
([A_1(p,y|x)\,,C(p,y|x)]_*)\Big |_{V_-}=0\,,
\eee
\be
\label{AC}
A_1(p,y|x) =\o(p,y|x)
-e^n \int_0^1 dt
\f{\p}{\p p^n}C(t^{-1}p,t y|x)\, +
\Big([A_1(p,y|x)\,,C(p,y|x)]_*\Big) \Big |_-\,,
\ee
where $\o(p,y|x)$ is the generating function for
HS gauge fields, i.e. an arbitrary
1-form such that $\tau_+ (\o (p,y|x))=0$.
The equation (\ref{AC}) determines
$A_1(p,y|x)$ perturbatively as an expansion in powers
of the HS gauge field $\o(p,y|x)$
and the Weyl 0-forms $C(p,y|x)$. The equation (\ref{DC}) is the
unfolded equation on $C(p,y|x)$.

Inserting (\ref{AC}) into (\ref{DAP})  one obtains a nonlinear
differential equation on $\o(p,y|x)$ of the form
\bee
\label{FDA1}
R^{mn}p_m y_n +R^m p_m &+&D^L\o (p,y|x) + \o(p,y|x)*\wedge \o (p,y|x)\nn\\
&=&
e^n\wedge e^m \f{\p^2}{\p p^m \p y^n} C^l (p,y|x) +O(\o,C)\,,
\eee
where $O(\o,C)$ contains at least one power of $C$ and is
at least bilinear in $\o$ and/or $C$. These terms are nonlinear
because both $C$ and $\o$ (which does not contain the
gravitational gauge fields) are treated as small perturbations.

The equations (\ref{FDA1}) and (\ref{DC}) provide nonlinear
unfolded off-shell constraints for all HS fields in flat space.
All nonlinear corrections can  be reconstructed order
by order by iterations of the equation (\ref{AC}). The consistency
is guaranteed by construction.

\section{Nonlinear unfolded equations for lower spins}
\label{field equations for lower spins}

Now let us discuss a possibility of putting off-shell systems
on-shell by imposing nontrivial field equations. At the linearized
level it is well-known \cite{LV,5d}
(see also  \cite{solv}) that to put a
system  on-shell it is enough to require the gauge fields and
field strengths to be traceless in the fiber indices. It is also
well-known \cite{diff} that it is not possible to extend the
on-shell formulation for massless spins $s> 2$ to the nonlinear level
unless a nonzero cosmological constant is introduced \cite{FV1},
for which case the problem was solved in \cite{more,non} (and references
therein). On the other hand, since the full off-shell system
admits consistent lower spin reductions and the latter admit
consistent nonlinear field equations like Yang-Mills and Einstein
equations, it is interesting to see how these equations look like
in terms of the unfolded formulation suggested in this paper. As
we show in this section the lower spin equations keep the standard
form in the unfolded formulation although formulated in terms of
0-forms $\BB$ in place the original 1-form gauge fields $A$.

\subsection{Yang-Mills theory}
\label{Yang-Mills}

Yang-Mills theory is described in terms of the gauge
1-form $A$ and
Weyl zero form $\BB$ having the form (\ref{bound}) with the background
gravitational fields $A_0$ (\ref{A0}) and $\BB_0$ (\ref{b0}) and
dynamical spin 1 fields $A_1=A_1(y|x)$ and $\BB_1 =p_n \BB_1^n (y)$
of zero and first order in $p_n$, respectively.
Here the fields $A_1=A_1(y|x)$ and $\BB_1 =p_n \BB_1^n (y)$ take
values in a Yang-Mills Lie algebra as well as the $p$-independent gauge
parameter $\epsilon (y|x)$.

The key observation is that, because of the vacuum part
$\BB_0$ (\ref{b0}), the spin one 0-form
$\BB_n$ transforms as Yang-Mills connection in the
$y$-space
\be
\delta \BB_n(y|x) = D^y_n \epsilon (y|x)\,,
\ee
where
\be
D^y_m \epsilon (y|x) =
\f{\partial}{\partial y^m} \epsilon (y|x)+ [\BB_m (y|x)\,, \epsilon (y|x)]\,.
\ee
As a result,
\be
\label{ycur}
\BB^{mn}(y|x) = \f{\partial}{\partial y_m} \BB^n (y|x)
-\f{\partial}{\partial y_n} \BB^m (y|x)
+[\BB^m (y|x) \,,\BB^n (y|x)]\,
\ee
behaves as $y$-space field strength and transforms
covariantly
\be
\delta \BB_{nm}(y|x) = [\BB_{nm} (y|x)\,, \epsilon (y|x)]\,.
\ee
The $x$-space Yang-Mills equations result from those in the
 $y$ space
\be
D_n^y \BB^{nm}(y|x) =0\,.
\ee
Although this conclusion may look like a miracle,
it is just a consequence of the identical form of the
Yang-Mills symmetries in the $x$-space and $y$-space.

Analogously to the case of scalar field it is not difficult to
write a general $Q$ closed Lagrangian for spin one
\be
\label{ymg}
\cl = \epsilon_{n_1\ldots n_d } e^{n_1}\wedge \ldots \wedge e^{n_d}
 \ell (\BB_{nm}, D^y_k \BB_{nm},\ldots )\Big|_{y=0}\,,
\ee
where $\ell (\BB_{nm}, D^y_k \BB_{nm},\ldots )$ is a
Lagrangian function of the $y$--field strength $\BB^{mn}$ and its
$y$-covariant derivatives taken at $y=0$, which is invariant
under the $y$--dependent Yang-Mills symmetry transformations.
In particular, the Lagrangian of the standard Yang-Mills theory
has the form
\be
\label{ym}
\cl = \epsilon_{n_1\ldots n_d } e^{n_1}\wedge \ldots \wedge e^{n_d}
 tr(\BB^{nm}(0|x) \BB_{nm}(0|x))\,,
\ee
where $tr$ is trace in a chosen matrix representation of the
Yang-Mills algebra.
This Lagrangian is manifestly gauge invariant under the gauge transformations
with the parameters $\epsilon(y|x)$ and therefore is $Q$ closed.
That it is equivalent to the standard Yang-Mills Lagrangian
follows from the fact that, taking into account that
the nonlinear term in (\ref{ycur}) is at least linear in $y$,
\be
\BB^{nm}(0|x)=\Big (\f{\partial}{\partial y_m} \BB^n (y|x)
-\f{\partial}{\partial y_n} \BB^m (y|x)\Big )\Big |_{y=0}=2 C^{n,m}(x)\,,
\ee
where $C^{n,m}(x)$ is Maxwell 0-form equal to the $x$--space
Yang-Mills field strength by virtue of the $p$--independent part
of (\ref{curls}).

\subsection{Gravity}
\label{Gravity}

To describe gravity one observes that the 0-form $\BB^{nm}(y|x)$
transforms like the metric tensor under the $y$--diffeomorphisms
(\ref{deb2}). Since it starts with the flat part
\be
\label{bper}
\BB^{nm}(y|x) = \half
\eta^{nm} +\BB^{nm}_1 (y|x)\,
\ee
 the matrix $\BB^{nm} (y|x)$ is (perturbatively)
invertible. The invariant conditions on $\BB^{nm}(y|x)$ have a form of
standard Einstein equations in the $y$ space
\be
\label{ye}
\car_{nm} (\BB(y|x))
=0 \,,\qquad \car_{nm} = \car_{nk,ml} \BB^{kl}\,,
\ee
where
$\car_{nm,kl}(\BB)$ is the Riemann tensor constructed from
$\BB^{nm}(y|x)$ treated as the metric tensor in the $y$--space.
Containing second derivatives, the $y$--space Einstein
equations (\ref{ye}) impose conditions on the components of
the $y$--space Riemann tensor $\BB^{nm}{}_{pq}
y^p y^q$ and higher order polynomials in $y$. The zero curvature
equations of the unfolding approach map these conditions to the
true Einstein equations on the frame $e^n$ where
$\BB^{nm}{}_{pq}$ is identified with the $x$--space Riemann
tensor by (\ref{curls}).

Note that
the expansion in powers of $y^n$ is a version of the normal
coordinate expansion. Also let us note that such a mechanism does
not work for second-order HS equations of massless fields because
the nontrivial components of the spin $s$ $0$-forms start from
order--$s$ derivatives of the HS potentials $\o(p,y|x)$. This is
in agreement with the expectation that HS massless fields do not
admit nontrivial interactions in flat Minkowski background
\cite{diff}. On the other hand, one can speculate that the problem
can be avoided even in flat space within the nonlocal approach
developed in \cite{nonlo}.

A general Lagrangian for the case of gravity can be written in
the form
\bee
\label{gract}
\cl &=& \epsilon_{n_1\ldots n_d } e^{n_1}\wedge \ldots \wedge e^{n_d}
\sqrt{(-1)^{d+1}\epsilon^{m_1\ldots m_d } \epsilon^{k_1\ldots k_d} \BB_{k_1 m_1}
\BB_{k_2 m_2}\ldots \BB_{k_d m_d}}\nn\\
&{}& \ell (\BB^{nm},\car_{nm,kl}, D^y_p \car_{nm,kl},\ldots )\Big |_{y=0}\,.
\eee
This Lagrangian is invariant under the
$y$-diffeomorphisms and (therefore) is  $Q$ closed provided that
all indices in
$ \ell (\BB^{nm},\car_{nm,kl}, D_p \car_{nm,kl},\ldots )$ are contracted
covariantly. Indeed, because all quantities in the
 Lagrangian (\ref{gract}) transform covariantly under $y$--diffeomorphisms
and because the Lagrangian is taken at $y=0$, only the global
$GL(d)$ part of $y$--space diffeomorphisms acts nontrivially
in (\ref{gract}). This $GL(d)$ part rotates the fiber indices,
leaving the Lagrangian invariant provided that all indices are
contracted covariantly. The only subtlety is that, although
the epsilon symbol $\epsilon^{m_1\ldots m_d }$ is only $SL(d)$
covariant, the combination of the epsilon symbols with upper and
lower indices  in (\ref{gract}) is $GL(d)$ covariant.

The Einstein action results from
\be
\ell = \BB^{nm}\car_{nm}\,
\ee
as is elementary to see in the gauge (\ref{vacb}) in which the
spin two fluctuational part $\BB^{nm}_1 (y|x)$ carries at least
two powers of $y$. In this gauge, the $y$--independent part
of $\BB^{nm} (y|x)$ is just the constant term in (\ref{bper}).
As a result, the square root in (\ref{gract}) becomes a
constant. Analogously to the Yang-Mills case,
the $y$--space Riemann tensor at $y=0$ is given by the
expression linear in the Riemann 0-form
$\BB^{nm}{}_{pq}$ identified with the $x$--space Riemann
tensor by (\ref{curls}).

\section{Contractions}
\label{contr}

The property that HS fields form infinite towers follows from
the structure of the star product algebra. However, because no
obstruction for unfolding
Bianchi identities with no dynamical equations imposed
can be expected, it should be possible
to  unfold any off-shell HS system in the background
gravitational and/or Yang-Mills fields.
In practice, this is achieved by replacing the
star product algebra with the commutative algebra of
functions of commuting variables $y^n$ and $p_n$
in the spin $s\neq 2$ sector, keeping  the Poisson algebra
in the spin two sector of connections linear in $p$.
 In other words, one considers the semidirect product
of the commutative associative algebra $H$ spanned by
the polynomials $P(p,y)$ of all powers in $p^n$ except
for linear polynomials (clearly polynomials linear in
$p$ never appear in a product of any two such functions).
Then one considers
a Lie algebra $t$ of linear in $p$ polynomials formed by
their Weyl commutators or, that is equivalent for
the linear functions in $p$ (i.e., vector fields),
by the Poisson brackets
\be
\label{pb}
 \{p_m,y^n\} = \delta_m^n
\ee resulting from the ``classical limit''     of (\ref{py}). This
gives rise to the spin two gauge fields. Then one defines the
action of $t$ on $H$ as a Poisson bracket \be a(f) = \{a,f\}\qquad
\forall a\in t\,,\quad f\in H \,. \ee Note that from this
definition it follows that $a(f)$ has the same power in $p_m$ as
$f(p,y)$, i.e. in this algebraic setup different spins do not mix.
The unfolded equations keep the same form (\ref{d2}) but with the
commutator replaced with the Poisson bracket with the spin two
connection $A(p,y|x)$ linear in $p$. In addition, one can
introduce the non-Abelian Yang-Mills structure by allowing the
spin one (i.e., $p$-independent) 1-form gauge potentials to be
matrices that allows
for a Yang-Mills covariantization by extending the algebra to the
semidirect sum of the HS commutative algebra with the spin two -- one
non-Abelian algebra. In this construction, off-shell HS fields are
independent and may carry any representations of the Yang-Mills
algebra. Moreover, in this setup it is possible to consider any
number of HS fields of any spin. For example: an off-shell
 system of gravity, Yang-Mills theory
with the Yang-Mills algebra $h$ and a spin three field in any
given representation of $h$ is possible. (That, of course, in no
way means that this system can be put on shell.)

There is also an intermediate truncation of the system (\ref{d2})
with the
star product replaced with the Poisson brackets (\ref{pb}).
All these off-shell formulations are formally consistent. Moreover,
we admit that they even should be equivalent by an appropriate
pseudolocal field redefinition analogous to the Seiberg-Witten
map \cite{SW} (more precisely, an off-shell system is expected to be
equivalent to its further contraction with the same spectrum of
fields).

To impose dynamical equations in these terms means to
restrict somehow the 1-form connection $A(p,y|x)$ and the 0-form
$\BB(p,y|x)$. Because the system (\ref{d2}) is overdetermined
it is not trivial to do it in a consistent way. Here is where the
role of $AdS$ background and star product structure becomes really
important. The original star product version of the system
seems to be fundamental because it is inherited  from
the fundamental system in $AdS_d$ of \cite{non} that admits
nontrivial field equations.

\section{Conclusion}

In this paper the general unfolded dynamics approach is extended
to the action level for off-shell unfolded systems.
Analogous construction for the on-shell unfolded systems
gives conserved charges. Both types of objects are
represented by cohomology of the derivation $Q$
of the $L_\infty$ algebra associated with the
unfolded system at hand.

Also we present a very simple unfolded form for off-shell
symmetric bosonic
HS fields of all spins in Minkowski space. The constraints
have a form of covariant constancy and zero curvature conditions
for 1-forms and 0-forms taking values in an appropriate
star product algebra. The form of these equations is very
suggestive. Their extension to fermionic systems seems to
be straightforward along the lines of \cite{0404124} and will be
given elsewhere.

The off-shell HS system in Minkowski space, considered in this
paper, results from the flat limit of the $AdS$ HS system of
\cite{non}. Details of this correspondence will be given in a
future publication. Let us just mention that the 0-form $\BB$ of
this paper is one of the covariantly constant $sp(2)$ generators
that underly the construction of \cite{non}.

An interesting problem for the future is to apply the general
scheme proposed in this paper to the construction of the action
principle in the $AdS$ HS gauge theory. Also it is interesting to
extend the obtained results to superstring. It is tempting to
speculate that the generators of $sp(2)$,
which play the key role both in the $AdS$ analysis of \cite{non}
and in the Minkowski analysis of this paper, should extend to the
Virasoro algebra via extension of a pair of oscillators $p^n$,
$y_n$ to the infinite set $p^n_i$, $y^n_i$ with $i=0,1,2\ldots$
{\it a la} the Moyal formulation of superstring \cite{bars}.

\section*{Acknowledgments}

I am grateful to  M.Grigoriev, E.Ivanov
and I.Tipunin for useful discussions.
The work  was supported in part by grants RFBR No.
05-02-17654, LSS No. 1578.2003-2 and INTAS No. 03-51-6346.

\newcounter{appendix}
\setcounter{appendix}{1}
\renewcommand{\theequation}{\Alph{appendix}.\arabic{equation}}

\section*{Appendix A.\\ Unfolded equations with manifest $x$--dependence}

The unfolded equations also make sense
when the differential $Q$ and/or
dynamical variables on which it acts contain the
manifest dependence on the coordinates $x^\un$ on the top of
that due to $W^\ga (x)$. In this case we set
\be
\label{qdifx}
Q=\partial + G^\ga (W,x)  \f{\p}{\p W^\ga}\,,
\ee
where the partial exterior differential
$
\partial =dx^\un \f{\p}{\partial x^\un }
$
takes into account only on the manifest dependence on $x,$ {\it i.e.}
\be
\partial \left( F(W(x),x)\right ) = \Big (dx^\un \f{\p}{\partial x^\un }
       F(W,x)\Big )\Big |_{W=W(x)}\,.
\ee
The condition $Q^2 =0$ takes the form
\be \label{BIx}\partial G^\ga (W,x)+ G^\gb (W,x)\wedge \f{\p
G^\ga (W,x)} {\p W^\gb}  =0\,.
\ee
The equation (\ref{unf1}) extends to
\be
\label{unf1x}
d F(W(x),x) =  Q (F(W(x),x)).
\ee
One can see that it remains invariant under the gauge transformations
(\ref{delw}). Note also
that for $W$--independent functions $F=F(x)$ the equation
(\ref{unf1x}) reduces to the identity $d F(x) = \p F(x)$.

\setcounter{appendix}{2}
\setcounter{equation}{0}
\renewcommand{\theequation}{\Alph{appendix}.\arabic{equation}}

\section*{Appendix B.\\ Dynamical content via $\s_-$ cohomology}
\label{taudynamcontent}

In this Appendix, we perform following \cite{sigma,BHS,s3,solv}
a very general analysis of equations
of motion of the form \be \widehat{D}_0\cc =0\ ,
\label{sfieldequs} \ee via a cohomological reformulation
  of the problem.
Here $\cc$ is a differential form of degree $p$ taking
values in a vector space $V$, that is an element of the complex $V\otimes
\Omega^p({\cal M}^d)$.
We require $V$ to admit a grading operator $G$ and
$\widehat{D}_0$ to
decompose as the sum
\be
\widehat{D}_0=\s_-+D_0+\s_+\,,\label{Dzero}
\ee
such that the following properties are true:
\begin{description}
  \item[(i)] The grading operator $G$ is diagonalizable
in the vector space $V$
and it possesses a spectrum bounded from below.
  \item[(ii)]
$$[G,D_0^L]=0\,,\qquad
[G,\s_-]=-\s_-\,.$$ The operator $\s_+$ is a sum of operators of
strictly positive grade.
  \item[(iii)] The operator $\s_-$ acts
vertically in the fibre $V$,  {\it i.e.} it does not act on
space-time coordinates. (In HS models, only the operator $D_0$
acts nontrivially on the space-time coordinates (differentiates).)
  \item[(iv)] The operator
$\widehat{D}_0$ (\ref{Dzero}) is nilpotent
\be
(\widehat{D}_0)^2=0\,.
\ee
and increases a form degree by one.
\end{description}

The graded decomposition of the nilpotency equation
gives the following identities \be
(\s_-)^2=0\,,\quad D_0\s_-+\s_- D_0=0\,,\quad
(D_0)^2+\s_+\s_-+\s_-\s_+ + D_0 \s_+ +\s_+ D_0
+(\s_+)^2=0\,.\quad \ee If $\s_+$ has definite grade $+1$,
the last relation is
equivalent to the three conditions $ (\s_+)^2=0\,,\quad D_0\s_+
+\s_+ D_0=0\,,\quad (D_0)^2+\s_+\s_-+\s_-\s_+ =0\,.$
\noindent An important property is the nilpotency of $\s_-$. The
key fact is that the analysis of Bianchi identities is
equivalent to the analysis of the cohomology of $\s_-$, that is
$$H(\s_-,V)\equiv \frac{Ker(\s_-)}{Im(\s_-)}\,.$$

The field equation (\ref{sfieldequs}) is
invariant under the gauge transformations
\be\d
\cc=\widehat{D}_0\varepsilon\,,\label{sgginv}\ee
since
$\widehat{D}_0$ is nilpotent by the hypothesis (iv). The
 gauge parameter $\varepsilon$
is a $(p-1)$-form. These gauge transformations contain both
differential gauge transformations (like linearized diffeomorphisms)
and Stueckelberg gauge symmetries (like linearized local Lorentz
transformations\footnote{Recall that, at the linearized level, the
metric tensor corresponds to the symmetric part $e_{\{\m\;m\}}$ of
the frame field. The antisymmetric part of the frame field
$e_{[\m\;m]}$ can be gauged away by fixing locally the Lorentz
symmetry, because it contains as many independent components as the
Lorentz gauge parameter $\varepsilon^{nm}$.}).

The following terminology will be used. By {\it dynamical field}, we
mean a field that is not expressed as derivatives of something else
by field equations ({\it e.g.} the frame field in gravity or a
frame-like HS 1-form field $e_{\m}^{ ~n_1 \ldots n_{s-1}}$). The
fields that are expressed by virtue of the field equations as
derivatives of the dynamical fields modulo Stueckelberg gauge
symmetries are referred to as {\it auxiliary fields} ({\it e.g.} the
Lorentz connection in gravity or its HS analogues $\o_{\m}^{ ~n_1
\ldots n_{s-1}, \, m_1 \ldots m_t}$ with $t>0$). A field that is
neither auxiliary nor pure gauge by Stueckelberg gauge symmetries is
said to be a {\it nontrivial dynamical field} ({\it e.g.}, the metric
tensor or the metric-like gauge fields of Fronsdal's
approach).\vspace{.2cm}

Let $\cc (x)$ be a section of the fiber bundle with space-time
(coordinate $x$) as the base manifold and
 $V\otimes \Omega^p({\cal M}^d)$ as the fibers,
that satisfies the equation (\ref{sfieldequs}).
Under the hypotheses (i)-(iv) the following
propositions are true \cite{sigma,BHS}:
\begin{itemize}
    \item[A.] Nontrivial dynamical fields $\cc$ are
     nonvanishing elements of $H^p(\s_-)$.
    \item[B.] Differential gauge symmetry parameters $\varepsilon$ are
    classified by $H^{p-1}(\s_-)$.
    \item[C.] Inequivalent differential field equations on the
     nontrivial dynamical fields, contained in $\widehat{D}_0\cc=0$, are in
one-to-one
    correspondence with elements of $H^{p+1}(\s_-)$.
\end{itemize}

\noindent \underline{Proof of A}: The first claim is almost obvious.
Indeed, let us decompose the field $\cc$ according to the grade $G$:
$$\cc=\sum_{n=0}\cc_n\,,\qquad G\cc_n=n\,\cc_n\,,
\qquad(n = 0,1,2,\ldots)\,.$$ The field equation (\ref{sfieldequs})
thus decomposes as \be \widehat{D}_0\cc|_{\,n-1}=
\s_-\cc_n+D_0 \cc_{n-1}+    \Big ( \s_+\sum_{m\leq
n-2}\cc_{m}\Big)\Big|_{n-1}=0\,. \label{decompsfieldequ}\ee By a
straightforward induction on $n=1,2,\ldots$, one can convince
oneself that all fields $\cc_n$ that contribute to the first term of
the right hand side of the equation (\ref{decompsfieldequ}) are
thereby expressed in terms of derivatives of lower grade ({\it i.e.}
$<n$) fields, hence they are auxiliary\footnote{Here we use the fact
that the operator $\s_-$ acts vertically (that is, it does not
differentiate space-time coordinates) thus giving rise to algebraic
conditions which express auxiliary fields via derivatives of the
other fields.}. As a result only fields annihilated by $\s_-$ are
not auxiliary. Taking into account the gauge transformation
(\ref{sgginv}) \be \d
\cc_n=\widehat{D}_0\varepsilon|_n=\s_-\varepsilon_{n+1}+
D_0\varepsilon_n+
\Big (\s_+\sum_{m\leq n-1} \varepsilon_{m} \Big )\Big |_{n}
\label{proofAB}\ee one observes that, due to the first term in this
transformation law, all components $\cc_n$ which are $\s_-$ exact,
{\it i.e.} which belong to the image of $\s_-$, are Stueckelberg and
they can be gauged away. Therefore, a nontrivial dynamical $p$-form
field in $\cc$ should belong to the quotient
$Ker(\s_-)/Im(\s_-)$.$\Box$ \vspace{.2cm}

\noindent For Einstein-Cartan's gravity, the Stueckelberg gauge
symmetry is the local Lorentz symmetry and indeed what distinguishes
the frame field  from the metric tensor is that the latter
belongs to the cohomology $H^1(\sigma_-)$ while the former
contains a $\s_-$ exact part. \vspace{.3cm}

\noindent \underline{Proof of B}: The proof follows the same lines
as the proof of A. The first step has already been performed in the
sense that (\ref{proofAB}) already told us that the parameters such
that $\s_-\varepsilon\neq 0$ are Stueckelberg and can be used to
completely gauge away trivial parts of the field $\cc$. Thus
differential parameters must be $\s_-$ closed. The only subtlety is
that one should make use of the fact that the gauge transformation
$\d_\varepsilon \cc=\widehat{D}_0\varepsilon$ are reducible. More
precisely, gauge parameters obeying the reducibility identity
\be\varepsilon=\widehat{D}_0\zeta\label{reducid}\ee are trivial in
the sense that they do not perform any gauge transformation,
$\d_{\varepsilon=\widehat{D}_0\zeta} \cc=0$. The second step of the
proof is a mere decomposition of the reducibility identity
(\ref{reducid}) in order to see that $\s_-$ exact parameters
correspond to reducible gauge transformations\footnote{ Note that
factoring out the $\s_-$ exact parameters accounts for algebraic
reducibility of gauge symmetries. The gauge parameters in
$H^{p-1}(\s_-)$ may still have differential reducibility analogous
to differential gauge symmetries for nontrivial dynamical fields.
}.
$\Box$ \vspace{.3cm}

\noindent \underline{Proof of C}: Given a nonnegative integer number
$n_0$, let us suppose that one has already obtained and analyzed
(\ref{sfieldequs})  in grades ranging from $n=0$ up to $n=n_0-1$.
Let us analyze (\ref{sfieldequs}) in grade $G$ equal to $n_0$ by
looking at the constraints imposed by the Bianchi identities.
Applying the operator $\widehat{D}_0$ on the covariant derivative
$\widehat{D}_0{\cal C}$ gives identically
 zero, which is the Bianchi identity
$(\widehat{D}_0)^2\cc=0$. Decomposing the latter Bianchi identity
gives, in grade equal to $n_0-1$, \be\label{Bidty}
(\widehat{D}_0)^2\cc|_{n_0-1}=\s_-\Big(\widehat{D}_0\cc|_{n_0}\Big)
+D_0\Big(\widehat{D}_0\cc|_{n_0-1}\Big)+\Big(\s_+ \sum_{m\leq
n_0-2} \widehat{D}_0\cc|_m\Big)\Big|_{n_0-1}=0\,.\ee By the
induction hypothesis, the equations $\widehat{D}_0\cc|_{m}=0$ with
$m\leq n_0 -1$ have already been imposed and analyzed.
Therefore (\ref{Bidty}) leads to
$$\s_-\Big(\widehat{D}_0\cc|_{n_0}\Big)=0\,.$$ In other words,
$\widehat{D}_0\cc|_{n_0}$ belongs to $ Ker(\s_-)$. Thus it can
contain a $\s_-$ exact part and a nontrivial cohomology part:
$$\widehat{D}_0\cc|_{n_0}=\s_-(E_{n_0+1})+F_{n_0}\,,\qquad F_{n_0}\in
H^{p+1}(\s_-)\,.$$ The exact part can be compensated by a field
redefinition of the component $\cc_{n_0+1}$ which was not treated
before (by the induction hypothesis). More precisely, if one
performs
$$\cc_{n_0+1}\rightarrow \cc^\prime_{n_0+1}:=\cc_{n_0+1}-E_{n_0+1}\,,$$
then one is left with $\widehat{D}_0\cc^\prime|_{n_0}=F_{n_0}$.
 The field equation (\ref{sfieldequs}) in grade $n_0$ is
$\widehat{D}_0\cc^\prime|_{n_0}=0$. This not only expresses the
auxiliary $p$-forms $\cc^\prime_{n_0+1}$ (that are not annihilated
by $\s_-$) in terms of derivatives of lower grade $p$-forms
$\cc_{k}$ ($k\leq n_0$), but also sets $F_{n_0}$ to zero.
If the cohomology is non-zero, this
imposes some $\cc_{n_0+1}$-independent conditions on the derivatives
of the fields $\cc_{k}$ with $k\leq n_0$, thus leading to
differential restrictions on the nontrivial dynamical fields.
Therefore, to each representative of $H^{p+1}(\s_-)$ corresponds a
differential field equation. $\Box$\vspace{.3cm}

Note that if $H^{p+1}(\s_-) =0$, the equation (\ref{sfieldequs})
contains only constraints which express auxiliary fields via
derivatives of the dynamical fields, imposing no restrictions on the
latter. If $D_0$ is a first order differential operator and if
$\s_+$ is at most a second order differential operator (which is
true in HS applications) then, if $H^{p+1}(\s_-)$ is nonzero in the
grade $k$ sector, the associated differential equations on a grade
$\ell$ dynamical field are of order $k+1-\ell$.

\end{document}